\documentclass{elsart}
\usepackage{graphicx,amssymb}
\newcommand{\be}{\begin{equation}}
\newcommand{\ee}{\end{equation}}
\newcommand{\bea}{\begin{eqnarray*}}
\newcommand{\bean}{\begin{eqnarray}}
\newcommand{\eea}{\end{eqnarray*}}
\newcommand{\eean}{\end{eqnarray}}
\newcommand{\beal}{\begin{large}\begin{eqnarray*}}
\newcommand{\eeal}{\end{eqnarray*}\end{large}}
\newcommand{\beanl}{\begin{large} \begin{eqnarray}} \newcommand{\eeanl}{\end{eqnarray}\end{large}}
\newcommand{\bel}{\begin{large}\begin{equation}}
\newcommand{\eel}{\end{equation}\end{large}}
\newcommand{\nn}{\nonumber}
\newcommand{\x}{{\rm x}}
\newcommand{\y}{{\rm y}}

\newcommand{\somma}{\sum_{n=-\infty}^{+\infty}}
\newcommand{\pl}{\frac{\pi}{L}}
\newcommand{\nor}{\frac{1}{\sqrt{2L}}}
\newcommand{\sq}{\sqrt{2}}

\newcommand{\0}{|0\rangle}
\newcommand{\1}{|\Omega^{(1)}\rangle}
\newcommand{\2}{|\Omega^{(2)}\rangle}

\newcommand {\psir}{\psi_R (0,\x)}

\begin{document}

\begin{flushright}
SMUHEP/04-04
\end{flushright}

\begin{frontmatter}

\title{Adjoint ${\rm QCD_{1+1}}$ in Light-cone Gauge, 
Quantized at Equal Time\thanksref{thanks1}}
\thanks[thanks1]{Work supported in part by the Department of Energy
under contract number DE-FG03-95ER40908.}

\author{E. Vianello}
\address{Department of Physics, Southern Methodist University,
Dallas, TX 75275}

\author{G. McCartor}
\address{Department of Physics, Southern Methodist University,
Dallas, TX 75275}

\begin{abstract}
$SU(2)$ gauge theory coupled to massless fermions in the adjoint representation  is quantized
in light-cone gauge
by imposing the equal-time canonical algebra.  The theory is defined on  a space-time cylinder 
with "twisted"  boundary conditions, periodic for one color component (the diagonal 3- component) 
and  antiperiodic for the other two.  The focus of the study is on the non-trivial vacuum structure 
and the fermion condensate. It is shown that the indefinite-metric quantization of free gauge 
bosons is not compatible with the residual gauge symmetry of the interacting theory. A suitable 
quantization of the unphysical modes of the gauge field is necessary in order to guarantee the 
consistency of the subsidiary condition and allow the quantum representation of the residual gauge 
symmetry of the classical Lagrangian: the 3-color component of the gauge field must be quantized in 
a space with an indefinite metric while the other two components require a positive-definite metric. 
The contribution of the latter to the  free  Hamiltonian becomes highly pathological  in this 
representation, but a larger portion of the interacting Hamiltonian can be diagonalized, thus  
allowing  perturbative calculations to be performed. The vacuum is evaluated through second order 
in perturbation theory and this result is used for an approximate determination of the fermion condensate.
\end{abstract}


%

\end{frontmatter}

\section{Introduction}
\label{sec:Introduction}

In this paper we study SU(2) gauge theory with the quarks in the adjoint representation. 
The model has been of interest since it possesses multiple ground states and a chiral 
condensate\cite{witten}\cite{paniak}\cite{smilga}\cite{kogan} \cite{lenz}
\cite{paper}.  We shall quantize at equal time with periodicity conditions on the space 
interval $-L\leq \x \leq L$. We shall use ``twisted" boundary conditions \cite{thooft} 
\cite{shifman}, periodic for one color component (the diagonal 3-component) and  
antiperiodic for the other two.  We note that these boundary conditions break the gauge
symmetry but they lead to worthwhile technical simplifications and have been used in past 
studies; our work here will find most in common with \cite{lenz}, where the authors used 
equal-time quantization and reduced to gauge independent degrees of freedom by hand,and 
\cite{paper}, where the authors used the light-cone gauge but quantized on the light-cone.

There are several reasons for wanting to study the light-cone gauge case using equal-time 
quantization.  In the case of the Schwinger model the light-cone gauge solution quantized 
at equal-time with periodicity conditions has more in common with the continuum solution 
than the solution quantized on the light-cone with periodicity conditions\cite{gmc}
\cite{eliana}.  In particular, the chiral condensate goes to zero at large $L$ in the case 
of light-cone quantization whereas it goes to the continuum value at large $L$ in the 
equal-time case.  Therefore, the authors of \cite{paper} were not able to make an estimate 
of the physical value of the condensate; we shall make such an estimate in the present paper.  
Also, the quantization at equal-time requires the use of Lagrange multiplier fields and an 
indefinite metric representation space, whereas, such fields are not required (or permitted) 
in the case of light-cone quantization with periodicity conditions and the representation space 
includes only physical states.  In the continuum case, whether quantized at equal-time or on 
the light-cone, the Lagrange multiplier fields are required\cite{nm} and the representation space must be 
of indefinite metric.  The properties of the Lagrange multiplier fields in the equal-time case with 
periodicity conditions are much like those of the continuum case  We expect these qualitative 
differences between the equal-time periodic case and the light-cone periodic case to hold for the 
nonabelian case as well and the results presented below will, to some extent, justify that 
expectation.

In the present paper we shall quantize in light-cone gauge at equal time with twisted boundary 
conditions through the use of Lagrange multiplier fields.  We explicitly construct the physical 
subspace and demonstrate that it is stable under time evolution.  We also construct the algebra 
of the Lagrange multiplier fields; that algebra is likely to be the same as the continuum case and 
it may be of use in attempts to construct a continuum solution, particularly if quantizing on the 
light-cone.  In setting up the quantization we encounter an unexpected difficulty: the system 
possesses a residual gauge symmetry at the classical level where the gauge matrix is given by
\be 
U(x)=e^{iN\pl (t+\x)\tau^3}
\ee
It is this residual gauge symmetry that leads to the multiple vacua which is something we want to 
study.  If this symmetry is not implemented at the quantum level the multiple vacua are not present.  
Since we have introduced unphysical degrees of freedom we would expect to have to quantize the 
components of the gauge field in indefinite metric in order to be able to consistently remove the 
unphysical states.  That is what is necessary in the case of the Schwinger model\cite{eliana}.  But 
here we find that if we quantize all the components of the gauge field in indefinite metric we cannot 
implement the residual gauge symmetry at the quantum level.  On the other hand, if we quantize all the 
components of the gauge field in positive metric, we cannot consistently remove the unphysical states.  
the solution, for the present case, is to quantize the periodic component of the gauge field in 
indefinite metric and the antiperiodic components of the gauge field in positive metric.  We show that 
that procedure allows us to implement the residual gauge symmetry and to consistently remove the 
unphysical states.  With the use of the mixed quantization scheme we then find that there are two 
possible vacua, in agreement with the findings in \cite{witten}\cite{smilga}\cite{lenz}\cite{paper}. 
The success of the mixed quantization procedure depends on the breaking of the gauge symmetry through 
the use of the twisted boundary conditions.  It is an open question as to how the quantization 
should be done in the continuum or in a case where the same periodicity conditions are imposed on all 
components of the gauge field.  The question may have some importance since the same issue arises in the 
light-cone quantization of standard QCD.

Once the quantization is set up we write the Hamiltonian in a form suitable for perturbative calculations.  
The existence of the condensate is a nonperturbative effect and neither the vacuum nor any other 
state can be calculated by a perturbative calculation in which the interaction is the perturbing 
operator.  But we find that we can find an ``unperturbed'' operator consisting of the kinetic energies 
plus a small part of the interaction which we can diagonalize in closed form.  The eigenstates of this 
``unperturbed'' operator contain all the singularities in the coupling constant and we can then perform 
standard perturbation calculations using the rest of the interaction as the perturbing operator.

We use the perturbative formalism to calculate the vacuum through second order.  We then use the vacuum 
to calculate the condensate through the same order.  We are able to find the exact dependence of the 
condensate on the parameters but have a constant (a pure number) for which we have only an expansion.  
We use Pad\'e approximants to estimate the value of this number in the limit $L \rightarrow \infty$.  
It would be interesting to compare our estimate with estimates obtained by other means but we currently 
know of no such calculations.

Our light-cone conventions are as follows:
\bea
&&\x^{-}=\frac{t-\x}{\sqrt{2}} \; \;\;\;\;\; ,\; \;\;\;\;\; \x^{+}=\frac{t+\x}{\sqrt{2}} \\ \\&&\partial_{-}=\frac{\partial}{\partial \x^{-}}=\frac{\partial_{0}-\partial_{1}}{\sqrt{2}}\;\;\;\; \; \;,\; \;\;\;\;\;\partial_{+}=\frac{\partial}{\partial \x^{+}}=\frac{\partial_{0}+\partial_{1}}{\sqrt{2}} \;.
\eea

\section{SU(2) Gauge Theory Coupled to Adjoint Fermions}

\subsection{Basics}

The lagrangian density for the theory is\footnote{The notation used here is similar to that of ref.\cite{paper}.}
\[
{\mathcal L} =-\frac{1}{2}\rm{Tr}(F_{\mu\nu}F^{\mu\nu})+\rm{Tr}(\bar{\Psi}\gamma^{\mu}D_{\mu}\Psi)
\]
where  \ \
\[F_{\mu\nu}=\partial_{\mu}A_{\nu}-\partial_{\nu}A_{\mu} + ig[A_{\mu},A_{\nu}] \quad , \quad D_{\mu}=\partial_{\mu}+ig[A_{\mu},\quad] \;.  \]
%
%
%
%
$A_{\mu}$ and $\Psi$ are matrices in the adjoint representation of $SU(2)$:
\[ A_{\mu}=A_{\mu}^{a}\tau^{a} \quad , \quad \Psi=\Psi^{a}\tau^{a} \qquad a=1,2,3
\]
where $ \tau^{a}=\frac{\sigma^{a}}{2}$
and $\sigma^{a}$ are the Pauli matrices, so that
\[
\left[\tau^a, \tau^b \right]=i\epsilon^{abc}\qquad ,
\qquad {\rm Tr} \left(\tau^a \tau^b \right)=\half \delta^{ab}\;.
\]
$\gamma^{0}$ and $\gamma^{1}$ are $2\times 2$ matrices satisfying the Dirac algebra
\[
\{\gamma^\mu, \gamma^\nu\} =2g^{\mu\nu}\,.
\]
We shall use the following representation
\[
\gamma^{0}=\left(\begin{array}{cc}0 & -i\\ i & 0\end{array}\right)\; \;, \; \;\gamma^{1}=\left(\begin{array}{cc}0 & i\\ i & 0\end{array}\right) \; .
\]
$\Psi^{a}$ is a 2-component Dirac field :
\[\Psi^{a}=   \left(\begin{array}{cc}\Psi^{a}_{R}\\  \Psi^{a}_{L}
\end{array}\right) \]
The lagrangian is invariant under the gauge transformation
\[
\Psi _{R/L}^\prime = U \Psi _{R/L} U^{-1}\; ,
\]
\[
A^\prime_{\mu} = UA_\mu U^{-1}+\frac{i}{g}\partial_\mu U U^{-1}
\]
where $U$ is a spacetime-dependent element of SU(2).\\
Note that $F_{\mu \nu}$ and $D_\mu$ transform covariantly under gauge
transformations:
\[
F_{\mu \nu}^\prime = U F_{\mu\nu}U^{-1} \qquad,
\qquad D_\mu ^\prime = U D_\mu U^{-1}\;.
\]
The equations of motion for the gauge fields are
\be\label{eqofmotion}
D_\mu F^{\mu \nu}=g J^\nu
\ee
where the fermion current $J^\mu$ is defined as
\be\label{currents}
J^\mu \equiv i{\bar \Psi}^b \gamma^\mu \Psi^a \left[ \tau^a, \tau^b \right] \; .
\ee
The conservation law  associated with the gauge invariance is
\[
\partial_\nu \left( J^\nu -i \left[ A_\mu, F^{\mu\nu }\right]   \right)=0 \; ,
\]
the fermion current being conserved in the covariant sense:
\[
D_\mu J^\mu =0 \; .
\]


We shall work in the light-cone gauge.  The light-cone gauge condition
\be \label{gauge}
  n^\mu A_\mu ^a=0 \quad \qquad {\rm with}\quad n=\frac{1}{\sqrt{2}}(1,-1)
\ee
or, equivalently,
\be
 A_ - ^a  \equiv \frac{1}{\sqrt{2}} (A_0 ^a -A_1 ^a )=0 \: ,
\ee
can be enforced by means of a Lagrange multiplier $\lambda(x)=\lambda^a (x)\tau^a$ , by adding
the gauge-fixing term \cite{bassetto}
\[
{\mathcal L}_{gf}=2{\rm Tr}\left(\lambda \, n^\mu A_\mu \right)
 \]
to the Lagrangian.
The theory defined by the Lagrangian
\[{\mathcal L}'={\mathcal L}+{\mathcal L}_{gf} \]
can be consistently quantized by means of Dirac's procedure\cite{dirac}. The gauge conditions
(\ref{gauge}) can be obtained as the Euler-Lagrange equations associated to the
fields $\lambda^a (x)$.

The quantum commutators corresponding to the classical Dirac's
brackets are
\[
[A_0^a (t,\x), (\pi^1)^b (t, \y)]= [A_1^a (t,\x), (\pi^1)^b (t, \y)]= \delta_{ab}(\x-\y)
\]
where $( \pi^1)^b=F_{01}^b$. We can see that the gauge constraint $A_- ^a=0 $ can be imposed in a strong sense while
$A_+^a$ satisfies
\[
[A_+ ^a (t,\x), F_{01}^b (t, \y)]= \sq \delta_{ab}(\x-\y) \;.
\]
This procedure introduces spurious degrees of freedom into the theory. The
Euler-Lagrange equations associated to the gauge fields
\be
D_\mu F^{\mu \nu}+\lambda n^{\nu}=g J^\nu
\ee
are not equivalent to eqs.(\ref{eqofmotion}) owing to the presence of the Lagrange multiplier. Equivalence with
the original theory can be recovered by imposing the subsidiary condition
$\lambda=0$. However, since the commutators of $\lambda$ with the other
fields are not zero, such
condition is incompatible with the quantization of the theory and
cannot be imposed in a strong sense. As in the standard Gupta-Bleuler quantization
of QED in the Feynman gauge, the subsidiary  condition will have to be imposed as a weak condition selecting
the physical subspace ${\mathcal V}_{phys}$ of the theory:
\be
|phys \rangle \in  {\mathcal V}_{phys} \Leftrightarrow
\langle phys | \lambda |phys \rangle =0
\ee
The stability of the physical subspace under time evolution is guaranteed
by the  fact that, as we shall see,  $\lambda$ satisfies a free-field equation of motion and has, therefore, a well defined decomposition into positive and negative frequency parts, so that one can equivalently state the subsidiary condition as
\be \label{subsidiary}
|phys \rangle \in {\mathcal V}_{phys}\Leftrightarrow
\lambda^{(+)} |phys \rangle =0
\ee
where $\lambda^{(+)}$ denotes the annihilation, or positive frequency, component
of the field $\lambda(x)$\,.


It is convenient to introduce the helicity basis \cite{paper}
\[\tau^{+}=\frac{\tau^{1}+i\,\tau^{2}}{\sqrt{2}}\qquad , \qquad \tau^{-}
=\frac{\tau^{1}-i\,\tau^{2}}{\sqrt{2}}
\]
These satisfy
\be \label{commtau}
[\tau^{+}, \tau^{-}]=\tau^{3} \qquad , \qquad [\tau^{3}, \tau^{\pm }]=\pm \tau^{\pm}
\ee
and
\be
\rm{Tr}(\tau^{+} \tau^{-})=\rm{Tr}(\tau^{3})^{2}=\frac{1}{2} \quad ,
\quad \rm{Tr}(\tau^{\pm})^{2}=\rm{Tr}(\tau^{3}\tau^{\pm})=0 \; .
\ee
With respect to this basis $A_{\mu}$ and $\Psi$ are decomposed as
\be
A_\mu = A_\mu ^3 \tau ^3 +A_\mu ^- \tau ^+ +A_\mu ^+ \tau ^-
\ee
where $A_\mu ^\pm \equiv \frac{1}{\sqrt{2}}(A_\mu ^1 \pm A_\mu ^2)$ ,
\be
\Psi _{ R/L}=
\phi_{R/L}\tau ^3+  \psi _{R/L}\tau ^+ +  \psi_{R/L} ^\dag \tau ^-
\ee
where  $\phi_{R/L} \equiv \Psi_{R/L}^3$ and $\psi_{R/L} \equiv \frac{1}{\sqrt 2}\left(\Psi ^1 _{R/L}- i \Psi ^2 _{R/L}\right)$\, ,
 $\psi^\dag _{R/L} \equiv \frac{1}{\sqrt 2}(\Psi ^1 _{R/L}+ i \Psi ^2 _{R/L})$\, .\\

We shall restrict  the space
variable to the interval $-L \leq \x \leq L$  and impose ``twisted"
boundary conditions: the fields $\psi_R$ and $\psi_L$ will
be taken to be antiperiodic; it will be convenient, however, to take
$\phi_R$ and $\phi_L$ to be periodic.  For
consistency, then, $A_\mu^\pm$ must be taken to be antiperiodic
while $A_\mu^3$ is periodic.

With the above definitions the Lagrangian density can be written as

\baselineskip15pt
\bean
\!\!\!{\mathcal L}&\!=&\! \!-\half (F_{01}^3 )^2 -\half F_{01}^+ F_{01}^- -\half F_{01}^- F_{01}^+
+\frac{i}{\sqrt{2}}\left[\phi_{R}\partial_{+}\phi_{R}+ \phi_{L}\partial_{-}\phi_{L}\right]
\nn \\ \nn \\
&&\! +\frac{i}{\sqrt{2}}\left[\psi_{R}^\dag \partial_{+}\psi_{R}+\psi_{R} \partial_{+}\psi_{R}^\dag +\psi_{L}^\dag \partial_{-}\psi_{L}+\psi_{L}\partial_{-}\psi_{L}^\dag  \right]\nn \\ \\
&&\! -{g\over\sqrt{2}}\left[A^3 _+  J^3 _R  +A^- _+  J^+ _R   +A_+ ^+  J^- _R +A^3 _-  J^3 _L  +
A^- _-  J^+ _L+ A_- ^+  J^- _L \right] \nn \\ \nn \\
&&+\lambda^3 A_- ^3+\lambda^+ A_- ^- +\lambda^- A_- ^+  \nn
\eean
where
\[
A_- ^{3,\pm} \equiv \frac{1}{\sqrt{2}}\left(A_0 ^{3,\pm}-A_1 ^{3, \pm}\right) \qquad , \qquad
A_+^{3,\pm} \equiv \frac{1}{\sqrt{2}}\left(A_0 ^{3,\pm}+A_1 ^{3, \pm}\right)
\]
and
\[
J^{3,\pm}_L =\frac{1}{\sqrt{2}}\left(J^{3,\pm}_0 + J^{3,\pm}_1\right) \qquad , \qquad
J^{3,\pm}_R =\frac{1}{\sqrt{2}}\left(J^{3,\pm}_0 -J^{3,\pm}_1\right)
\]
The equations of motion for the gauge fields  take the form
\bean
&&\partial_- F^3 +J^3 _R =0 \\
&&\partial_+ F^3 +i g (F^+ A^- - F^- A^+) -J^3 _L +\lambda^3 =0 \\
&&\partial_- F^- +J^- _R =0\\
&&\partial_+ F^- +i g (F^- A^3-F^3 A^-  ) -J^- _L +\lambda^- =0 \\
&&\partial_- F^+ +J^+ _R =0 \\
&&\partial_+ F^+ +i g (F^3 A^+-F^+ A^3  ) -J^+ _L +\lambda^+ =0\\
&& A_- =0
\eean
where $A^{3,\pm}\equiv A^{3,\pm}_+ $ and $F^{3,\pm}\equiv
F^{3,\pm}_{01}$, which is the only non-vanishing component of the
antisymmetric tensor $F^{3,\pm}_{\mu\nu}$ in two dimensions. The
condition $A_-=0$ implies
\[
F^{3,\pm}=\partial_0 A^{3,\pm} _1 -\partial_1 A^{3,\pm}_0 =\partial_- A^{3,\pm}
\]
From the expression of the energy-momentum tensor
\[
\Theta^{\mu \nu }=\frac{\partial\mathcal{L}}{\partial(\partial_{\mu}\varphi_{\alpha})}\partial^{\nu}\varphi_{\alpha}-{\mathcal L}g^{\mu \nu}
\]
one obtains the canonical Hamiltonian
\[
 P^{0}\equiv H=\int_{-L}^{L}d\x\, \Theta^{00}(x) .
\]
where
\bea
\Theta^{00}&= &F^3 \partial_0 A_1 ^3 +F^- \partial_0 A_1 ^+ +F^+ \partial_0 A_1 ^-
+\frac{i}{2}\left(  \phi_R \partial_0 \phi_R +  \psi^\dag _R \partial_0 \psi_R  + \psi _R \partial_0 \psi^\dag _R  \right) \\
&& +\frac{i}{2}\left(  \phi_L \partial_0 \phi_L + \psi^\dag _L \partial_0 \psi_L +\psi _L \partial_0 \psi^\dag _L \right)-{\mathcal L}
\eea
With some manipulations and  using the constraint $A_- =0$ one gets
\bea
H&=&\int_{-L}^{L} d\x \, \Bigg\{ \half  \left(F^3 \right)^2 +F^+ F^-  -
\frac{1}{\sqrt{2}}\left(\partial_1 F^3 A^3 +\partial_1 F^+ A^- +\partial_1 F^- A^+  \right) \\
&&+\frac{i}{2}\left( \phi_{L}\partial_{1}\phi_{L} +\psi_{L}^\dag \partial_{1}\psi_{L}+\psi_{L}\partial_{1}\psi_{L}^\dag -
\phi_{R}\partial_{1}\phi_{R}-\psi_{R}^\dag  \partial_{1}\psi_{R}-\psi_{R} \partial_{1}\psi_{R}^\dag  \right)\\
&&+ g  \left(A^3   J^3 _R  +A^-  J^+ _R   +A^+  J^- _R \right) \Bigg\}\;. 
\eea

\subsection{Quantization of the Fermi field}

Our treatment of the fermi field is standard.  The Fock representation for the fermionic degrees of freedom at $t=0$ is
obtained by Fourier expanding $\Psi_{R/L}(0,\x)$ .  We have
\begin{eqnarray}
\phi _R(0,\x) &=& {1 \over \sqrt {2L}} \sum_{N=1}^\infty
\left(r_N e^{ik _N \x} + r{_N ^\dag} e^{-ik_N \x} \right)
+{\stackrel {\;o} {\phi}}_R  \label{phiR}\\
\psi _R(0,\x)&=& {1 \over \sqrt {2L}} \sum_{n={1\over2}}^\infty
\left(b_n e^{ik _n \x} + d{_n ^\dag} e^{-ik_n \x} \right)
\label{psiR} \\
\phi _L(0,\x) &=& {1 \over  \sqrt {2L}} \sum_{N=1}^\infty
\left(\rho_N e^{-ik_N \x} + \rho{_N ^\dag}
e^{ik_N \x} \right) +{\stackrel {\;o} {\phi}}_L \label{phiL}\\
\psi _L(0,\x) &=& {1 \over \sqrt {2L}} \sum_{n={1\over2}}^\infty
\left(\beta_n e^{-ik _n \x} + \delta{_n ^\dagger}
e^{ik_n \x} \right) \; ,
\end{eqnarray}
where ${\stackrel {\;o} {\phi}}_{R/L}$ are the zero modes of $\phi_{R/L}$.
The lower-case (upper-case) indices run over positive
half-odd integers (integers) and $k_n= n \pi /L$, $k_N= N \pi /L$ .

The canonical anti-commutation relations for the Fermi fields are
\bea
\left\{\phi_{R} (0,\x),\phi_{R}(0,y)\right\}
&=&  \delta _{P}(x-y)  \\
\left\{\phi_{L} (0,\x),\phi_{L}(0,y)\right\}
&=&  \delta_{P} (x-y) \; ,
\eea
where $\delta_{P}$ denotes the periodic delta function, which can be expanded
in the interval $[-L, +L]$ as
 \[
\delta_P (\x-\y)=\frac{1}{2L}\sum_{N=-\infty}^\infty e^{i \frac{\pi}{L}N(\x-\y)} \;,
\]
and
\bean
\left\{\psi_{R} (0,\x),\psi^\dag_{R}(0,y)\right\}
&=&  \delta _{A}(x-y)  \label{commpsiR}\\
\left\{\psi_{L} (0,\x),\psi_{L}^\dag (0,y)\right\}
&=&  \delta_{A} (x-y) \label{commpsiL}\;,
\eean
where $\delta_{A}$ denotes the anti-periodic delta function
\[
 \delta_A (\x-\y)=\frac{1}{2L}\somma e^{i \frac{\pi}{L}n(\x-\y)} \;.
\]
All the other anti-commutators vanish.\\
These induce the following algebra for the Fourier modes:
\be
\{\rho{^\dagger _N}, \rho_M \} = \{r{^\dagger _N}, r_M \}
= \delta_{N,M}
\label{rhccrs}
\end{equation}
\begin{equation}
\{ b{_n ^\dagger}, b_m \} = \{ d{_n^\dagger}, d_m \}
= \{ \beta{_n ^\dagger},\beta _m \}
= \{\delta {_n^\dagger},\delta_m\} = \delta _{n, m}
\label{lhccrs}
\end{equation}
\begin{equation}
\{{\stackrel {\;o} {\phi}}_R,{\stackrel {\;o} {\phi}}_R \} =
\{{\stackrel {\;o} {\phi}}_L,{\stackrel {\;o} {\phi}}_L \} =
{1\over2L}\; ,
\label{zmccrs}
\end{equation}
all other anti-commutators vanishing. \\
The fermionic Fock space is generated in the usual way by the action of the
creation operators on a vacuum state $|0\rangle$.

We must define the currents with a regularization procedure that is consistent with 
gauge invariance.  We shall use the gauge invariant point splitting procedure~\cite{paper} 
and define the currents as
\[
\hat{J} _R(0, \x) \equiv -\lim_{\epsilon \rightarrow 0}
\Psi^a _R(0,\x+\epsilon)
 \Psi^b _R (0, \x) \, \big[ e^{ ig \int_\x^{\x + \epsilon} A_1 \cdot d\x}\,\tau^a\,
e^{- ig \int_\x^{\x + \epsilon} A_1 \cdot d\x}, \tau^b \big]
\]
This definition gives the expected result that
\bean
&&\hat{J}_R ^3 = \tilde{J}_R ^3 +\frac{g}{\sqrt{2}\pi}A^3 _1 \nn \\
&&\hat{J}_R ^+ = \tilde{J}_R ^+ +\frac{g}{\sqrt{2}\pi}A^+ _1 \label{Rcurr}\\
&&\hat{J}_R ^- = \tilde{J}_R ^- + \frac{g}{\sqrt{2}\pi}A^-  _1\; . \nn
\eean
and
\bean
&&\hat{J}_L ^3 = \tilde{J}_L ^3 -\frac{g}{\sqrt{2}\pi}A^3 _1\nn \\
&&\hat{J}_L ^+ = \tilde{J}_L ^+ -\frac{g}{\sqrt{2}\pi}A^+ _1\\
&&\hat{J}_L ^- = \tilde{J}_L ^-  - \frac{g}{\sqrt{2}\pi}A^-  _1\; .\nn
\eean
Here, the tilde operators, $\tilde{J}$, are the normal ordered, but un-gauge-corrected,
currents:
\bea
\tilde{J}_R ^3 (0, \x)=\sqrt{2}\, \textbf{:}\psi_R ^\dag (0, \x)\psi_R (0,\x) \textbf{:}=
 \frac{1}{\sqrt{2}L}\sum_{m,n=\half}^{\infty}
\bigg(d_n b_m e^{i(k_n +k_m )\x}
-d_m ^\dag d_n e^{i(k_n -k_m )\x}\qquad \quad \\
+b_n ^\dag b_m e^{-i(k_n -k_m )\x}+
b_n ^\dag d_m ^\dag e^{-i(k_n +k_m )\x}\bigg) \; .
\eea
\bea
\tilde{J}^+ _R (0,\x)&=&\frac{1}{2L}\sum_{N=0}^{\infty}\sum_{n=\half}^{\infty}
\Big( r_N d_n e^{i(N+n)\pl}- b_n ^\dag r_Ne^{i(N-n)\pl}\\ &&\hspace{40pt}+
r_N ^\dag d_n e^{-i(N-n)\pl}+r_N ^\dag b_n ^\dag e^{-i(N+n)\pl}\Big)
\eea
\bea
\tilde{J}^- _R (0,\x)&=&\frac{1}{2L}\sum_{N=0}^{\infty}\sum_{n=\half}^{\infty}
\Big(b_n r_N  e^{i(N+n)\pl \x}-r_N ^\dag b_n  e^{-i(N-n)\pl \x}\\
&&\hspace{60pt}
+d_n^ \dag  r_N e^{i(N-n)\pl \x}+d_n^ \dag r_N ^\dag  e^{-i(N+n)\pl \x}\Big)
\eea
where we have set
\be \label{zeromode}
{\stackrel {\;o} {\phi}}_R =\frac{r_{0}+r_{0}^\dag}{\sqrt{2L}}\;.
\ee
With similar expressions for the $\tilde{J}_L$ currents.

We must also regularize the kinetic energy in a gauge invariant manner and again we shall use 
gauge invariant point splitting.  We define
\bea
&& \!\!\!\!\!\!\!\!\left[\rm{Tr} (i \Psi_R(0,\x) \partial_1 \Psi_R(0,\x)\right]_{reg} \equiv\\
&&\qquad \equiv \lim_{\epsilon \rightarrow 0}\left \{\rm{Tr}\!\left(i e^{ig\int_\x ^{\x+\epsilon} A_1 (0,\y)d\y} \Psi_R(0, \x+\epsilon) e^{-ig\int_\x ^{\x+\epsilon} A_1 (0,\y) d\y}
\partial_1 \Psi_R (0,\x)\right)
-\rm{v.e.v.}\right \}
\eea
Which leads to
\be\label{Rkinetic}
\left[\rm{Tr} (i \Psi_R(0,\x) \partial_1 \Psi_R(0,\x)\right]_{reg} =
\frac{i}{2} \, \textbf{:}\,\psi_R ^\dag \buildrel \leftrightarrow \over
\partial_1 \psi_R  \textbf{: }+\frac{i}{2}\,\textbf{:}\,\phi_R \partial_1 \phi_R  \textbf{: }
+\frac{g^2}{4\pi}\left(2 A^+ _1  A^- _1 +(A^3 _1 )^2\right)\;.
\ee
Analogously, 
\be\label{Lkinetic}
\left[\rm{Tr} (i \Psi_L(0,\x) \partial_1 \Psi_L(0,\x)\right]_{reg} =
\frac{i}{2} \, \textbf{:}\,\psi_L ^\dag \buildrel \leftrightarrow \over
\partial_1 \psi_L \textbf{: }+\frac{i}{2}\,\textbf{:}\,\phi_R \partial_1 \phi_L \textbf{: }
-\frac{g^2}{4\pi}\left(2 A^+ _1  A^- _1 +(A^3 _1 )^2\right)\; .
\ee

It will be convenient to perform our analysis in the bosonized basis.  To that end we shall
need the Fourier components of the tilde currents.  We write
\be\label{JR3}
\tilde{J}_R ^3 (0, \x)=\frac{1}{\sq L}
\sum_{N=1}^{\infty}\left(C_N ^3 e^{ik_N \x}+{C_N ^3 }^\dag e^{-ik_N \x} \right)+\frac{C_0 ^3}{\sq L}
\ee
where
\[
C_N^3 =\sum_{n={1 \over 2}}^\infty \left(b^\dagger_n b_{N+n}
-d^\dagger_n d_{N+n}\right)
-\sum_{n={1 \over 2}}^{N-{1 \over 2}} b_n d_{N-n}
\]
and
\begin{equation}
C^3_0 = \sum_n (b _n ^\dagger b_n - d_n ^\dagger d_n ) \; .
\end{equation}
Also
\be \label{JL3}
\tilde{J}_L ^3 (0, \x)=\frac{1}{\sq L}
\sum_{N=1}^{\infty}\left(D_N ^3 e^{-ik_N \x}+{D_N ^3 }^\dag e^{ik_N \x} \right)
+\frac{D_0 ^3}{\sq L}
\ee
where
\[D_N^3 = \sum_{n={1 \over 2}}^\infty \left(\beta^\dagger_n \beta_{N+n}
-\delta^\dagger_n \delta_{N+n}\right)
-\sum_{n={1 \over 2}}^{N-{1 \over 2}} \beta_n \delta_{N-n}
\]
and
\begin{equation}
D^3_0 = \sum_n (\beta _n ^\dagger \beta_n - \delta_n ^\dagger \delta_n ) \; .
\end{equation}
For the $\pm$ currents we have
\be \label{JRpm}
\tilde{J}_R ^\pm (0, \x)=\frac{1}{\sq L}
\sum_{n=\half}^{\infty}\left(C_n ^\pm e^{ik_n \x}+{C_n ^\mp }^\dag e^{-ik_n \x} \right)
\ee
where
\bean
C_n ^+ &=& \sum_{M=0}^\infty r^\dagger_M d_{n+M}
-\sum_{m={1 \over 2}}^\infty b^\dagger_m r_{n+m}
-\sum_{m={1 \over 2}}^{n} d_m r_{n-m} \label{Cn+}\\
C_n ^- &=& \sum_{m={1 \over 2}}^\infty d^\dagger_m r_{n+m}
-\sum_{M=0}^\infty r^\dagger_M b_{M+n}
-\sum_{m={1 \over 2}} ^{n} r_{n-m} b_m \label{Cn-}\; .
\eean
Similar expressions can be found for the operators $D_n ^\pm$ such that
\be\label{JLpm}
\tilde{J}_L ^\pm (0, \x)=\frac{1}{\sq L}
\sum_{n=\half}^{\infty}\left(D_n ^\pm e^{-ik_n \x}+{D_n ^\mp }^\dag e^{ik_n \x} \right)
\ee
Using  the fundamental anti-commutators (\ref{rhccrs}),
(\ref{lhccrs}) and  ( \ref{zmccrs}) one can  verify that these
operators satisfy the commutation relations\footnote{Note that these relations do not hold in this form if the zero mode of $\phi_{R/L}$ is discarded, as in ref. \cite{paper}}\cite{goddard}
\begin{eqnarray}
\left [ C^3_N , C^3_M         \right ] &=& N \delta_{N,-M}\label{commc3} \\
\left [ C_n^{\pm} , C_m^{\pm} \right ] &=& 0 \label{commcpm}\\
\left [ C^3_N , C_m^{\pm}     \right ] &=& \pm C_{N+m}^{\pm}
\label{commc3cpm}\\
\left [ C^+_n , C^-_m  \right ] &=& C^3_{n+m} + n \delta_{n,-m}\; . \label{commcpcm}
\end{eqnarray}
where we have defined
\[
C^3_{-N}\equiv (C^3_N)^\dag \ , \qquad C_{-n}^{\pm}\equiv (C_n^{\mp})^\dag
\]
 The algebra satisfied by the $D$s is of course identical.

Following  \cite{nakawaki} we define
\bea
\varphi_{R}^{(+)}(0,\x)&=&-\sum_{N=1}^{\infty}\ \frac{1}{N}{C_N^3}e^{ik_N \x} \\
\varphi_{R}^{(-)}(0,\x)&=&\sum_{N=1}^{\infty}\frac{1}{N}{(C_N^3)^\dag}e^{-ik_N\x} \\
\varphi_{L}^{(+)}(0,\x)&=&-\sum_{N=1}^{\infty} \frac{1}{N}{D_N^3}e^{-ik_N\x} \\
\varphi_{L}^{(-)}(0,\x)&=&\sum_{N=1}^{\infty}\ \frac{1}{N}{(D_N^3)^\dag}e^{ik_N\x}
\eea
and
\be \label{sigmaRL}
\sigma_{R/L}(0, \x)=\sqrt{2L}e^{\varphi_{R/L}^{(-)}(0,\x)}\psi_{R/L}(0,\x)e^{\varphi_{R/L}^{(+)}(0,\x)}\;.
\ee
The following relations hold \cite{nakawaki}
\be \label{sigmasigma+R}
\sigma_{R}^{+}(0,\x)\sigma_{R}(0,\x)=\sigma_{R}(0,\x)\sigma_{R}^{+}(0,\x)=1\; ,
\ee
\be \label{sigmasigma+L}
\sigma_{L}^{+}(0,\x)\sigma_{L}(0,\x)=\sigma_{L}(0,\x)\sigma_{L}^{+}(0,\x)=1\; ,
\ee
\be \{\sigma_{R}(x),\sigma_{L}(y)\}=\{\sigma_{R}(x),\sigma^{+}_{L}(y)\}=0 \; ,
\ee
\be
[C^3_0\, , \, \sigma_{L}]=[D^3_0\, , \, \sigma_{R}]=0
\ee
\be
[C^3_0\, , \, \sigma_{R}]=-\sigma_{R}\quad, \qquad [D^3_0\, , \, \sigma_{L}]=-\sigma_{L}
\ee
\be \label{1.28}
[C_N^3 ,\sigma_{R/L}]=[D_N^3,\sigma_{R/L}]=0\;.
\ee
The action of the spurions, $\sigma_{R/L}\equiv \sigma_{R/L}(0,0)$, on the vacuum is given
as follows:
\[
|M,N\rangle = \sigma_L^M \sigma_R^N |0\rangle \; ,
\]
where
\[ \sigma_{R/L}^{-N}\equiv (\sigma_{R/L}^\dag)^{N}\; .
\]
and for $M,N>0$ ,
\bea
&&|M,N\rangle =\delta_{M-\half}^\dag \cdots \delta_{\half}^\dag \,
 d_{N-\half}^\dag \cdots d_{\half}^\dag |0\rangle \\
&&|-M,N\rangle = \beta_{M-\half}^\dag \cdots \beta_{\half}^\dag \,
 d_{N-\half}^\dag \cdots d_{\half}^\dag |0\rangle \\
&&|M,-N\rangle = \delta_{M-\half}^\dag \cdots \delta_{\half}^\dag \,
 b_{N-\half}^\dag \cdots b_{\half}^\dag |0\rangle \\
&&|-M,-N\rangle = \beta_{M-\half}^\dag \cdots \beta_{\half}^\dag \,
 b_{N-\half}^\dag \cdots b_{\half}^\dag |0\rangle \; .
\eea
It is easy to see that the states  $|M,N\rangle $ are eigenstates of $C_0^3$ and $D_0^3$ :
\bea
C_0^3|M,N\rangle &=&-N|M,N\rangle\\
D_0^3|M,N\rangle &=&-M|M,N\rangle \; .
\eea
One can verify that, for any $P>0$
\[
C_P^3|M,N\rangle =0 \quad , \qquad D_P^3|M,N\rangle =0
\]
and since
\[
[C_0^3, {C_P^3}^\dag ]=[D_0^3, {C_P^3}^\dag ] =0 \\
\]
\[
[C_0^3, {D_P^3}^\dag ]=[D_0^3, {D_P^3}^\dag ]=0
\]
the action of ${C_P^3}^\dag$ and ${D_P^3}^\dag$ does not modify the eigenvalues
of $C_0^3$ and $D_0^3$.\\
It can be shown  \cite{uhlenbrock} that the fermion Fock space ${\mathcal {F}}$, generated
by the action of the creation operators $b_n^\dag$ , $d_n^\dag$ , $\beta_n^\dag$ , $\delta_n^\dag$
on the vacuum $|0\rangle$,
can be decomposed as an infinite  direct sum of irreducible representations of the
bosonic algebra satisfied by the operators $C^3_P$  and $D^3_P$ ($P\neq 0$), each
representation corresponding to an eigenspace of $C_0^3$ and $D_0^3$ . More
explicitly we have
\[
{\mathcal{F}}=\oplus_{M,N}{\mathcal{F}}_{MN} \qquad M,N=0,\pm1,\pm 2,\ldots
\]
where ${\mathcal{F}}_{MN}$  is the Fock space generated by applying products of the operators
${C_P^3}^\dag$ and ${D_P^3}^\dag$ to the vacuum $|M,N\rangle $ and
\[
 \forall \, |\Phi_{MN}\rangle \in {\mathcal{F}}_{MN}\; : \quad C_0^3|\Phi_{MN}\rangle =-N|\Phi_{MN}\rangle  \quad , \qquad D_0^3|\Phi_{MN}\rangle =-M|\Phi_{MN}\rangle  \;.
\]
The free fermion hamiltonian
\bean
H_{\psi}&=&\frac{i}{2} \int_{-L}^{L}d\x \left( \psi_L^\dag (0,\x)  \buildrel \leftrightarrow \over
\partial_1
\psi_L (0,\x)-\psi_R^\dag (0,\x)  \buildrel \leftrightarrow \over
\partial_1 \psi_R (0,\x) \right) \nn \\
&=&\sum_{n=\half}^{\infty} k_n (\beta_n^\dag \beta_n + \delta_n^\dag \delta_n
+b_n^\dag b_n + d_n^\dag d_n )
\eean
and the momentum operator
\bean
P_{\psi}&=&\frac{i}{2} \int_{-L}^{L}d\x \left( \psi_L^\dag (0,\x)  \buildrel \leftrightarrow \over
\partial_1 \psi_L (0,\x)+\psi_R^\dag (0,\x)  \buildrel \leftrightarrow \over
\partial_1 \psi_R (0,\x) \right) \nn \\
&=&\sum_{n=\half}^{\infty} k_n (\beta_n^\dag \beta_n + \delta_n^\dag \delta_n
-b_n^\dag b_n - d_n^\dag d_n) \label{p1}
\eean
can be expressed in terms of the boson operators by means of the Kronig identities:
\bean
&&H=\frac{\pi}{2L}\Big((C_0^3)^2+ (D_0^3)^2 \Big)+\frac{\pi}{L}\sum_{N=1}^\infty
\left({C_N^3}^\dag  C_N^3 +{D_N^3}^\dag  D_N^3 \right)
\; \\
&& P_{1}=\frac{\pi}{2L}\Big((D_0^3)^2- (C_0^3)^2 \Big)+\pl \sum_{N=1}^\infty
\left({D_N^3}^\dag  D_N^3 -{C_N^3}^\dag  C_N^3\right)
\label{p2}\;.
\eean
Finally, using
\[ \sigma_{R/L}(0,\x)=e^{-iP_{\psi}\x} \sigma_{R/L} e^{P_{\psi}\x}
\]
it is easy to see that
\bea
\sigma_R (0,\x)&=&e^{ \frac{i\pi}{2L}C_0^3\x }\sigma_R e^{ \frac{i\pi}{2L}C_0^3\x }\\
\sigma_L (0,\x)&=&e^{ -\frac{i\pi}{2L}D_0^3\x }\sigma_R e^{ -\frac{i\pi}{2L}D_0^3\x }
\eea
and  the operators $\psi_{R/L}$  at $t=0$ can then be written in terms of bosonic operators
as
\bean
\psi_R (0,\x)&=& \nor \, e^{-\varphi_R^{(-)} (0,\x)}e^{ \frac{i\pi}{2L}C_0^3\x }\sigma_R e^{ \frac{i\pi}{2L}C_0^3\x }e^{-\varphi_R^{(+)} (0,\x)} \label{psiRbos}\\
\psi_L (0,\x)&=& \nor \, e^{-\varphi_L^{(-)} (0,\x)}e^{- \frac{i\pi}{2L}D_0^3\x }\sigma_L e^{- \frac{i\pi}{2L}D_0^3\x }e^{-\varphi_L^{(+)} (0,\x)}\; .\label{psiLbos}
\eean

\subsection{the Hamiltonian and the Subsidiary Condition}

In this subsection we shall give the regularized quantum Hamiltonian and shall discuss the
subsidiary condition.  We shall show that the physical subspace is dynamically stable and 
shall discuss some properties of the physical states.

Using the regularized expressions (\ref{Rcurr}) and
(\ref{Rkinetic}) for the currents and the fermion kinetic terms we
obtain the regularized quantum hamiltonian \bean \hat
{H}&=&\int_{-L}^{L} d\x \, \Bigg\{ \half  \left(F^3 \right)^2 +F^+
F^-  -
\frac{1}{\sqrt{2}}\left(\partial_1 F^3 A^3 +\partial_1 F^+ A^- +\partial_1 F^- A^+  \right) \nn \\
&&+\frac{i}{2}\left({\bf :}  \phi_{L}\partial_{1}\phi_{L}{\bf :} +{\bf :} \psi_{L}^\dag \buildrel \leftrightarrow \over
\partial_{1}\psi_{L}{\bf :} +-{\bf :}
\phi_{R}\partial_{1}\phi_{R}{\bf :} -{\bf :} \psi_{R}^\dag  \buildrel \leftrightarrow \over
\partial_{1}\psi_{R}{\bf :}  \right)\nn \\
&&+ g  \left(A^3   J^3 _R  +A^-  J^+ _R   +A^+  J^- _R \right)
+ \frac{g^2}{4\pi}\left[ (A^3)^2 + 2A^+ A^- \right] \Bigg\}
\label{hamiltonian}
\eean
where the products of gauge fields will also have to be
defined.

Starting from the Fourier expansions of $A^a _1= \frac{1}{\sqrt {2}}A^a  \,$ and $F^a$ at $t=0$ in the space interval
$[-L,L]$ with the chosen boundary conditions :
\bea
A^3 _1(0,\x)&=&\nor \sum_{N} a^3 _N e^{-ik_N \x}\\
A^{1,2} _1(0,\x)&=&\nor \sum_{n} a^{1,2} _n e^{-ik_n \x}\\
F^3 (0,\x)&=&\nor \sum_{N}b^3 _N e^{-ik_N \x}\\
F^{1,2} (0,\x)&=&\nor \sum_{n}b^{1,2} _n e^{-ik_n \x}
\eea
and using $a^3 _{-N }= {a^3 _N}^ \dag$ , $b^3 _{-N} = {b^3 _N}^ \dag$ , $a^\pm _{-n}= {a_n ^1}^\dag
\pm i {a_n^2}^\dag = {a^\mp _n}^\dag$ , $b^\pm _{-n}={b^\mp _n}^\dag$ , we can write
\bean
A^3 (0,\x)&=&\frac{1}{\sqrt{L}}\,
\sum_{N=1}^{\infty}\left( a^3 _N e^{-ik_N \x}+{a^3 _N }^\dag e^{ik_N \x}\right) +
\frac{1}{\sqrt{L}}\, a_0 ^3 \\
A^\pm (0,\x)&=&\frac{1}{\sqrt{L}} \,\sum_{n=-\infty}^{\infty}a_n^\pm e^{-ik_n\pl\x}=\frac{1}{\sqrt{L}} \,
\sum_{n=\frac{1}{2}}^{\infty}\left( a^\pm _N e^{-ik_n \x}+{a^\mp _n}^\dag e^{ik_n \x}\right) \label{Apm}\\
F^3 (0,\x)&=& \nor \,
\sum_{N=1}^{\infty}\left( b^3 _N e^{-ik_N \x}+{b^3 _N }^\dag e^{ik_N \x}\right)
+ \nor \, b_0 ^3 \\
F^\pm (0,\x)&=&\nor \,\sum_{n=-\infty}^{\infty} b^\pm _n e^{-ik_n \x}=\nor \,\sum_{n=\frac{1}{2}}^{\infty}\left( b^\pm _n e^{-ik_n \x}+{b^\mp _n }^\dag e^{ik_n \x}\right) \label{Fpm}\; .
\eean
From the commutation relations
\[
[ A^a (0,\x), F^b(0,\y)]=i \sqrt{2}\delta^{ab }\delta(\x-\y)
\]
we see that we must have
\bea
&&[ A^\pm (0,\x), F^\mp (0,\y)]=i \sqrt{2}\delta(\x-\y)\\
&&[ A^3 (0,\x), F^3 (0,\y)]=i \sqrt{2}\delta(\x-\y)
\eea
and
\be\label{commab}
[a^3 _M, {b^3 _N}^\dag]=i\delta_{MN}
\quad, \quad
[a^\pm _m, {b^\pm _n}^\dag]=i\delta_{mn}
\quad , \qquad
\ee
all the other commutators vanishing.  We shall postpone the discussion of implementing this 
algebra in a representation space until after we have discussed the subsidiary condition.  For 
that discussion we need the regularized hamiltonian $\hat{H}$
\bean
\hat{H}&=& H^{0}_F +\half \left(b_0 ^3\right)^2 +\frac{g^2}{2\pi}\left(a_0 ^3\right)^2+
g \sqrt{\frac{2}{L}}    C_0^3 a_0 ^3 \nn \\
&&\!\! \!+\sum_{N=1}^{\infty}\bigg\{ {b^3 _N}^\dag b^3 _N
+i k_N {a^3_N}^\dag b^3 _N -i k_N { b^3 _N}^\dag
 a^3_N +\frac{g^2}{\pi}{a^3_N}^\dag a^3_N +
g\sqrt{\frac{2}{L}}\left(C^3 _N a^3 _N +{C^3 _N}^\dag {a^3 _N}^\dag\right)\!\! \bigg\}\nn\\
&&\!\!\!+ \sum_{n=\half}^{\infty}\bigg\{ {b^+ _n}^\dag b^+ _n +{b^- _n}^\dag b^- _n
+i k_n {a^+_n}^\dag b^+ _n -i k_n { b^- _n}^\dag a^-_n
+i k_n {a^-_n}^\dag b^- _n -i k_n { b^+ _n}^\dag a^+_n   \nn \\
&&\!\!\!\quad+\frac{g^2}{\pi}\left({a^+_n}^\dag a^+_n + {a^-_n}^\dag a^-_n \right)
+g\sqrt{\frac{2}{L}}\left( C_n ^+ a_n^- + C_n ^- a_n^+ + {C_n ^+}^\dag {a_n^-}^\dag
+ {C_n ^-}^\dag {a_n^+}^\dag \right)\!\!\bigg\}\!
\eean
where ${H}^0_F$ is the free fermion Hamiltonian
\be \label{HF}
{H}^0_F=\sum_{n=\half}^{\infty} k_n \left(\beta_n^\dag \beta_n + \delta_n^\dag \delta_n
+b_n^\dag b_n + d_n^\dag d_n \right)+\sum_{N=0}^{\infty}\left(\rho_N^\dag \rho_N +r_N^\dag r_N \right)
\ee
which, as we have seen, can also be expressed as
\be \label{HFbos}
H_F^0=
\frac{\pi}{2L}\Big((C_0^3)^2+ (D_0^3)^2 \Big)+\frac{\pi}{L}\sum_{N=1}^\infty
\left({C_N^3}^\dag  C_N^3 +{D_N^3}^\dag  D_N^3 \right)
+\sum_{N=0}^{\infty}\left(\rho_N^\dag \rho_N +r_N^\dag r_N \right) .
\ee 

The Lagrange multipliers $\lambda^3$, $\lambda^\pm$ are given in terms of the
other fields by
\bea
\lambda^{3}&=&-\sqrt{2}\partial_1 F^3 -igF^+ A^- +ig F^- A^+ +\sqrt{2}g\tilde{J}^3 _{0}\\
\lambda^{+}&=&-\sqrt{2}\partial_1 F^+ -igF^3 A^+ +ig F^+ A^3 +\sqrt{2}g\tilde{J}^+ _{0}\\
\lambda^{-}&=&-\sqrt{2}\partial_1 F^--igF^- A^3 +ig F^3 A^- +\sqrt{2}g\tilde{J}^- _{0}\; .
\eea
Consider the Fourier expansion
\bea
\lambda^3 (0,\x)&=&\nor \sum_{N=-\infty}^{\infty} \lambda ^3 _N e^{-ik_N \x}\; ,
\\
\lambda^{\pm}(0,\x)&=&\nor \sum_{n=-\infty}^{\infty}\lambda_n^\pm e^{-ik_n \x}\; .
\eea
Note that  from   $\lambda^3=(\lambda^3)^\dag $ and
 $\lambda^\pm=(\lambda^\mp)^\dag $ it follows that
$\lambda^3_{-N}=(\lambda^3_N)^\dag$ and
 $\lambda^\pm_{-n}=(\lambda^\mp_n)^\dag$.
We want to show that the time evolution of the Lagrange multipliers in the Heisenberg
picture is that of free fields satisfying the simple equation $\partial_- \lambda=0$.
In order to see this let us evaluate their commutators with the Hamiltonian.\\
$
[\hat{H} , \lambda^3 (0,\y)]
$
consists of the following terms:

\vspace{-40pt}
\baselineskip36pt
\bea
&1.&ig\int_{-L}^{L}d\x [F^+ (0,\x) F^- (0,\x)\,,\, F^- (0,\y) A^+ (0,\y) -F^+ (0,\x)A^- (0,\y)]\\
&2.& -\frac{ig}{\sq}\int_{-L}^{L}d\x \,[ \, \partial_1 F^+ (0,\x) A^- (0,\x)+\partial_1 F^- (0,\x) A^+ (0,\x)\, , \,
\\
&& \hspace{90pt}
F^- (0,\y) A^+ (0,\y) -F^+ (0,\y)A^- (0,\y)\, ]\\
&3.&\int_{-L}^{L}d\x \,[\, \partial_1 F^3 (0,\x)A^3  (0,\x) \,, \, \partial_1 F^3  (0,\y) \,] \\
&4.& g\pl\sum_{N=1}^{\infty}\, \left[\,{C_N ^3}^\dag C_N ^3+ {D_N ^3}^\dag D_N ^3\, , \, \tilde{J}^3_R (0,\y)+
\tilde{J}^3_L(0,\y)\, \right]\\
&&- g\sq \int_{-L}^{L}d\x \, \left[ \,A^3(0,\x)\tilde{J}_R^3 (0,\x)\,,\, \partial_1 F^3(0,\y)\, \right] \\
&5.&
g^2 \! \int_{-L}^{L}\! d\x \left[A^3(0,\x)\tilde{J}_R^3 (0,\x)\,,\tilde{J}_R^3 (0,\y)\right]\!
-\frac{g^2}{2\sq \pi}\int_{-L}^{L}\! d\x \left[(A^3)^2(0,\x) ,  \partial_1 F^3(0,\y)\right]
\\ &6.&
g^2 \int_{-L}^{L}d\x \, \left[\, A^- (0,\x) \tilde{J}^+_R(0,\x)\, , \, -iF^+ (0,\y) A^- (0,\y)+\tilde{J}^3_R (0,\y)\, \right]
\\ &7.&
g^2 \int_{-L}^{L}d\x \,\left[\,A^+ (0,\x) \tilde{J}^-_R(0,\x)\, , \, iF^- (0,\y) A^+ (0,\y)+\tilde{J}^3_R (0,\y)\, \right]
\\ &8.&
\frac{ig^3}{2\pi} \int_{-L}^{L}d\x \, \left[\,A^+ (0,\x) A^- (0,\x), F^- (0,\y)A^+ (0,\y)-F^+ (0,\y)A^- (0,\y)\, \right]
\eea
We have
\baselineskip24pt
\bea
1.&=&g\sq  \int_{-L}^{L}d\x \left(-F^+ (0,\y) F^- (0,\x) \delta_A (\x-\y)+
F^+ (0,\x) F^- (0,\y) \delta (\x-\y)\right)=0 \\ \\
2. &=&g \int_{-L}^{L}d\x \left(-\partial_{\x}F^+ (0,\x)A^-(0,\y)\delta_A (\x-\y)+F^+ (0,\y)A^-(0,\x)\partial_{\x}\delta_A (\x-\y)\right)\\
& &+g \int_{-L}^{L}d\x \left(\partial_{\x}F^- (0,\x)A^+(0,\y)\delta(\x-\y)-F^- (0,\y)A^+(0,\x)\partial_{\x}\delta(\x-\y)\right)\\
& &=g\, \partial_{\y}\Big(F^- (0,\y)A^+(0,\y)-F^+ (0,\y)A^- (0,\y)\Big)\\ \\
3. &=& i\sq \int_{-L}^{L}d\x \,\partial_{\x}F^3 (0,\x)\partial_{\y}\delta_P(\x-\y)=
i\sq\, \partial_{y}^2 F^3 (0,\y)\\ \\
4. &=&\frac{g}{2L}\sum_{N=1}^{\infty}\left( -k_N C_N^3 e^{ik_N \x}+
k_N {C_N^3}^\dag e^{-ik_N \x} -k_N D_N^3 e^{-ik_N \x}+
k_N {D_N^3}^\dag e^{ik_N \x}\right)+\\
&&-2ig\int_{-L}^{L}d\x \tilde{J}^3(0,\x)\partial_{\y}\delta_P (\x-\y)\\
&=&ig\partial_{\y}\tilde{J}^3_R(0,\y)-ig\partial_{\y}\tilde{J}^3_L(0,\y)-2ig\partial_{\y}\tilde{J}^3_R(0,\y)
=-ig\, \partial_{\y}\Big(\tilde{J}^3_R(0,\y)+\tilde{J}^3_L(0,\y)\Big)\\ \\
5. &=& \int_{-L}^{L}d\x \left[ \frac{g^2}{2L^2}A^3 (0,\x)\sum_{N=1}^{\infty}N \left(e^{ik_N(\x-\y)}-e^{-ik_N(\x-\y)}\right)-\frac{ig^2}{\pi}A^3(0,\x)\partial_{\y}\delta_P (\x-\y)\right]\\
&=& \int_{-L}^{L}d\x \left[ \frac{ig^2}{\pi}A^3(0,\x)\partial_{\y}\delta_P (\x-\y)-\frac{ig^2}{\pi}A^3(0,\x)\partial_{\y}\delta_P (\x-\y)\right]=0\\ \\
6. &=&\sq g^2 \tilde{J}^+_R(0,\y) A^- (0,\y)-\frac{g^2}{2L^2}\int_{-L}^{L}d\x A^- (0,\x) \sum_{n=-\infty}^{\infty}
\sum_{N=-\infty}^{\infty}C^+ _{n+N} e^{ik_n \x}e^{ik_N \y}\\
&=&\sq g^2 \tilde{J}^+_R(0,\y) A^- (0,\y)-\frac{g^2}{2L^2}\int_{-L}^{L}d\x A^- (0,\x)\sum_{n=-\infty}^{\infty}C^+ _{n} e^{ik_n \x}\sum_{N=-\infty}^{\infty}e^{-ik_N (\x-\y)}\\
&=&\sq g^2 \tilde{J}^+_R(0,\y) A^- (0,\y)-g^2\sq\int_{-L}^{L}d\x A^- (0,\x)
\tilde{J}^+_R (0,\x)\delta_P (\x-\y)=0\\ \\
7. &=&\sq g^2 \tilde{J}^-_R(0,\y) A^+ (0,\y)-g^2\sq\int_{-L}^{L}d\x A^+ (0,\x)
\tilde{J}^-_R (0,\x)\delta_P (\x-\y)=0 \\ \\
8. &=& \frac{g^2}{\sq\pi}\int_{-L}^{L}d\x \Big(-A^- (0,\x)A^+ (0,\y)\delta_A (\x-\y)+
A^+ (0,\x)A^- (0,\y)\delta_A (\x-\y)\Big)=0
\eea
\baselineskip30pt
so that
\[
[\hat{H} , \lambda^3 (0,\y)]=-i\partial_{\y} \lambda^3 (0,\y)
\]
and
\[
[\hat{H} , \lambda^3_N]=\nor
\int_{-L}^{L}d\y \, e^{ik_N \y}[\hat{H} , \lambda^3 (0,\y)]=-k_N \lambda^3_N \;.
\]
The time evolution of $\lambda^3$ is given by
\[
e^{i\hat{H}t}\lambda^3 (0,\x)e^{-i\hat{H}t}=\nor \sum_{N=1}^{\infty}\Big( \lambda ^3 _N e^{-ik_N (t+\x)}+ {\lambda ^3 _N }^\dag e^{ik_N (t+\x)}+\lambda_0 ^3\Big)
\]
Let us consider $[\hat{H} , \lambda^+ (0,\y)]$ . A similar calculation gives
\bea
\int_{-L}^{L} d\x \, &\! \textbf{\Big[}&\! \half  \left(F^3 (x)\right)^2 +F^+ (x)F^-  (x) -
\frac{1}{\sqrt{2}}\Big\{\partial_1 F^3 (x) A^3 (x) +\partial_1 F^+ (x)A^- (x)\\
&&+\partial_1 F^- (x)A^+ (x)  \Big\}
+ \frac{g^2}{4\pi}\left\{(A^3)^2 (x)+ 2A^+ (x)A^- (x)\right\}\, , \,
-\sqrt{2}\partial_1 F^+ (y)\\
&&-igF^3 (y) A^+ (y)
+ig F^+ (y) A^3 (y) \,\textbf{ \Big]}_{x_0 =y_0 =0}=\\
&=& -i\partial_{\y} \Big\{ -\sqrt{2}\partial_1 F^+ (0,\y)-igF^3 (0,\y) A^+ (0,\y)+ig
 F^+ (0,\y)A^3 (0,\y) \Big\}\\
&& -\frac{ig^2}{\pi}\partial_{\y}A^+ (0,\y)
\eea
and
\[
g^2 \int_{-L}^{L}d\x \Big\{[A^3 (0,\x) \tilde{J}^3_R(0,\x)\, , \, \tilde{J}^+ _R (0,\y)]+i [A^-  (0,\x)\tilde{J}^+_R (0,\x)\, , \, A^3 (0,\y)F^+ (0,\y) ]\Big\}=0
\]
while
\bea
 &&g^2 \int_{-L}^{L}d\x  \Big\{A^+ (0,\x)\left[\tilde{J}^- _R (0,\x)\, , \,\tilde{J}^+ _R (0,\y)\right]
-i  \tilde{J}^3 _R (0,\x)[A^3 (0,\x)\, , \,
F^3 (0,\y)] A^+ (0,\y) \Big\}\quad  \; \\
&&=g^2 \int_{-L}^{L}d\x  \Big\{\frac{A^+ (0,\x)}{2L^2}\sum_{m,n=-\infty}^\infty e^{ik_n \x}e^{ik_m \y}
[C_n ^-\, , \, C_m ^+] +\sq \tilde{J}^3 _R (0,\x)A^+ (0,\y) \delta (\x-\y) \Big\}\quad\\
&&=-\frac{g^2}{2L^2} \int_{-L}^{L}d\x A^+ (0,\x) \sum_{m,n=-\infty}^\infty e^{ik_n \x}e^{ik_m \y}\left(C^3_{m+n}+m \delta_{m,-n}\right)+g^2 \sq \tilde{J}^3 _R (0,\y)A^+ (0,\y) \\
&&=-\frac{g^2}{L} \int_{-L}^{L}d\x A^+ (0,\x)  \Big\{ \sum_{N=-\infty}^\infty C^3_{N}e^{ik_N \x}
\delta(\x-\y)+\frac{iL}{\pi}\partial_{\x}\delta(\x-\y)\Big\}\hspace{95pt}\\
&&\quad+g^2 \sq \tilde{J}^3 _R (0,\y)A^+ (0,\y) \hspace{290pt}\\
&&=\frac{ig^2}{\pi}\partial_{\y}A^+ (0,\y)\hspace{330pt}
\eea
and
\[
-g \sq  \int_{-L}^{L}d\x  \tilde{J}^+ _R (0,\x) [A^- (0,\x)\, , \, \partial_{\y} F^+ (0,\y)]=
-2ig\partial_{\y}\tilde{J}^+ _R (0,\y)
\]
so that
\bea
[\hat{H} , \lambda^+ (0,\y)]&=&-i\partial_{\y} \Big\{ -\sqrt{2}\partial_1 F^+ (0,\y)-igF^3 (0,\y) A^+ (0,\y)+ig F^+ (0,\y) A^3 (0,\y) \Big\}\\
&&-2ig\partial_{\y}\tilde{J}^+ _R (0,\y) +g \left[ H_F \, , \, \tilde{J}^+_R (0,\y)+  \tilde{J}^+_L (0,\y)\right]\; .
\eea
Let us evaluate $\left[ H_F \, , \, \tilde{J}^+_R (0,\y)\right]$ and $\left[ H_F \, , \, \tilde{J}^+_L (0,\y)\right]$ . We have
\bea
\left[ H_F \, , \,  C_n ^+\right] = \Big[ \sum_{m=\half}^{\infty}k_m (b_m^\dag b_m+d_m^\dag d_m)
+\sum_{M=1}^{\infty}k_M r_M^\dag r_M \; , \,  \sum_{N=0}^\infty r^\dagger_N d_{n+N}\\
\quad -\sum_{j={1 \over 2}}^\infty b^\dagger_j r_{n+j}
-\sum_{j={1 \over 2}}^{n} d_j r_{n-j} \Big]
\eea
and, using the relation
\[
[AB\, , \, CD]= A \{B\, ,C\}D-AC \{B\, , D\}+\{A\, ,  C\} DB-C\{A\, ,  D\} B
\]
we get, for positive $n$,
\bea
\left[ H_F \, , \,  C_n ^+\right] &=&
- \sum_{m,j=\half}^{\infty} k_m b_m^\dag r_{n+j}\delta_{mj}
- \sum_{m=\half}^{\infty}\sum_{N=0}^\infty k_m r^\dagger_N d_m \delta_{m, \, n+N}\\
&&- \sum_{m=\half}^{\infty}\sum_{j={1 \over 2}}^{n}k_m r_{n-j}d_m \delta_{mj}
+\sum_{M=1}^{\infty}  \sum_{N=0}^\infty k_M r_M^\dag  d_{n+N}\delta_{MN}\\
&&+\sum_{M=1}^{\infty}
\sum_{j={1 \over 2}}^\infty k_M b^\dagger_j r_M \delta_{M,\, n+j}
+\sum_{M=1}^{\infty} \sum_{j={1 \over 2}}^{n}k_M d_j  r_M \delta_{M,\, n-j}\\
&=&- \sum_{j=\half}^{\infty}( k_j -k_{n+j}) b_j^\dag r_{n+j}
- \sum_{N=0}^\infty (k_{n+N}-k_N ) r^\dagger_N d_{n+N}\\
&&- \sum_{j={1 \over 2}}^{n}(k_j +k_{n-j})d_j r_{n-j}\\
&=&-k_n C^+_n \;. \label{commutatorofHandC}
\eea
Analogously one obtains
\bea
&&\left[ H_F \, , \,  (C_n ^-)^\dag \right]=k_n (C^-_n)^\dag \\
&&\left[ H_F \, , \,  (D_n ^+) \right]=-k_n D^+_n \\
&&\left[ H_F \, , \,  (D_n ^-)^\dag \right]=k_n (D^-_n)^\dag \; .
\eea
Therefore we have
\[
\left[ H_F \, , \, \tilde{J}^+_R (0,\y)\right]=\frac{1}{\sq L}
\sum_{n=\half}^{\infty}\left(-k_nC_n ^+ e^{ik_n \y}+k_n{C_n ^- }^\dag e^{-ik_n \y} \right)
=i\partial_{\y}\tilde{J}^+_R (0,\y)
\]
and
\[
\left[ H_F \, , \, \tilde{J}^+_L (0,\y)\right]=
\frac{1}{\sq L}
\sum_{n=\half}^{\infty}\left(-k_n D_n ^+ e^{-ik_n \x}+k_n(D_n ^- )^\dag e^{ik_n \x} \right)
=-i\partial_{\y}\tilde{J}^+_L (0,\y) \;.
\]
Finally we can write
\bea
[\hat{H} , \lambda^+ (0,\y)]&=&-i\partial_{\y} \Big\{ -\sqrt{2}\partial_1 F^+ (0,\y)-igF^3 (0,\y) A^+ (0,\y)+ig F^+ (0,\y) A^3 (0,\y) \Big\}\\
&&-ig\partial_{\y}\left(\tilde{J}^+ _R (0,\y) +  \tilde{J}^+_L (0,\y)\right)\\
&=&-i\partial_{\y} \lambda^+ (0,\y)
\eea
and, obviously,
\bea
[\hat{H} , \lambda^- (0,\y)]=-i\partial_{\y} \lambda^- (0,\y)
\eea
so that
\bea
[\hat{H} , \lambda^\pm_n]=\nor \int_{-L}^{L}d\y e^{ik_n\y}[\hat{H} , \lambda^\pm (0,\y)]=
-k_n\lambda_n^\pm \;.
\eea
As a consequence we have
\[\lambda^\pm (t,\x)=
e^{i\hat{H}t}\lambda^\pm (0,\x)e^{-i\hat{H}t}
=\nor \sum_{n=\half}^{\infty}\Big( \lambda ^\pm _n e^{-ik_n (t+x)}
+ (\lambda ^\pm _n )^\dag e^{ik_n (t+x)}\Big)
\]
\baselineskip24pt
We have thus shown that
the Heisenberg field $\lambda(t, \x)$ has a free-field decomposition into positive
and negative frequency components, which is fundamental for a consistent quantization
of the theory. This result guarantees that the decomposition of $\lambda$ into
Fock creation and annihilation operators and the definition of the physical subspace
by means of the subsidiary condition are stable under time evolution. 

Another important result is
\[ [\hat{H}, \lambda_0^3 ]=0\,.
\]
Indeed, the zero mode of $\lambda^3$ is a conserved charge.
In order to satisfy the subsidiary condition,
we shall require that its physical eigenstates have zero eigenvalue.

To further investigate the structure of the physical subspace let us
consider the algebra of the Lagrange multipliers.
Using the canonical commutation relations we get

\bea
[\lambda^3 (0,\x),\lambda^+ (0,\y)]&=&-2g F^+ (0,\y) \partial_{\x} \delta_P (x-y)
-2gF^+ (0,\x) \partial_{\y}\delta_A (\x-\y) \\
&&+ i\sq g^2 F^+ (0,\x) A^3 (0,\y) \delta_A (x-y)\\
&&-i\sq g^2 F^3 (0,\y) A^+ (0,\x) \delta_A (x-y)\\
&& +g^2 [\tilde{J}_R ^3  (0,\x)+ \tilde{J}^3_L (0,\x)\, , \tilde{J}_R ^+ (0,\y)+ \tilde{J}^+_L (0,\y)]
\eea
and
\bea
[\lambda^3 _N, \lambda^+ _m] &=&\frac{1}{2L} \int _{-L}^{L} d\x \, e^{ik_N \x}
 \int _{-L}^{L}d\y \, e^{ik_m \y}\; [\lambda^3 (0,\x),\lambda^+ (0,\y)]\\
&=&\frac{ g}{\sq L} \int _{-L}^{L} d\x \, e^{i(k_N +k_m)\x}
\bigg[ i(k_N+k_m)\sq F^+ (0,\x)+igF^+ (0,\x)A^3 (0,\x)\\
&& \qquad \qquad  \qquad \qquad-ig F^3 (0,\x)A^+ (0,\x)\bigg]+ \frac{g^2}{L} [ {C_{-N} ^3} + D_N^3  ,
C_{-m} ^+  + D_m^+ ]\\
&=&\frac{ g}{\sq L} \int _{-L}^{L} d\x \, e^{i(k_N +k_m)\x}
\bigg[ -\sq \partial_1 F^+ (0,\x)+igF^+ (0,\x)A^3 (0,\x) \\
&&\qquad \qquad \qquad  -ig F^3 (0,\x)A^+ (0,\x)\bigg]+ \frac{g^2}{L}\left( C^+_{-N-m} + D_{N+m}^+  \right) \;
\eea
where  (\ref{JR3}),  (\ref{JL3}), (\ref{JRpm}), (\ref{JLpm}) and (\ref{commc3cpm})
were used.
Finally one has
\[
[\lambda^3 _N, \lambda^+ _m]=\frac{g}{\sqrt{L}}\; \lambda^+ _{N+m}
\]
and, analogously,
\[
[\lambda^3 _N, \lambda^- _m]=-\frac{g}{\sqrt{L}}\; \lambda^- _{N+m} \; .
\]
In particular, for $N=0$
\[
[\lambda^3 _0, \lambda^\pm _m]=\pm \frac{g}{\sqrt{L}}\; \lambda^\pm _{m} \; .
\]
which shows that $\lambda^\pm$ are charged fields. This result has the important consequence that the subsidiary conditions involving
$\lambda^\pm$ are identically satisfied for states with zero eigenvalue of the charge $\lambda_0^3$ :
\[
\langle phys |\lambda^\pm _n |phys \rangle =\pm\frac{g}{\sqrt{L}}
\langle phys |[\lambda_0^3 \, , \,\lambda^\pm _n ]|phys \rangle = 0
\]
as long as
\[ \lambda_0 ^3 |phys \rangle =0 \; .
\]
Therefore we only need to require that physical states satisfy the conditions
\bea
&& \lambda_N^3 |phys \rangle =0 \quad {\rm for \ N>0}\;, \\
&&\lambda_0^3 \,|phys \rangle =0 \; .
\eea
One can also show that
\bea
[\lambda_n ^+ , \lambda_m ^- ]&=&\frac{ g}{\sq L} \int _{-L}^{L} d\x \, e^{i(k_n +k_m)\x}
\bigg[ -\sq \partial_1 F^3 (0,\x)+igF^- (0,\x)A^+ (0,\x) \\
&&\qquad \qquad \qquad  -ig F^+ (0,\x)A^- (0,\x)\bigg]+ \frac{g^2}{L} [ {C_{-n} ^+} + D_n^+  , C_{-m} ^-  + D_m^- ]
\eea
which, using \ $\left [ C^+_{-n} , C^-_{-m}   \right ] = C^3_{-n-m} - n \delta_{n,-m}$
\  and\   $\left [ D^+_n , D^-_m   \right ] =D^3_{n+m} + n \delta_{n,-m}$,  gives
\[
[\lambda^+ _n, \lambda^- _m]=\frac{g}{\sqrt{L}}\; \lambda^3 _{n+m}
\]
and

\baselineskip30pt
\bea
[\lambda^3(0,\x) \, , \, \lambda^3(0,\y)] &=& g^2[\tilde{J}^3_R(0,\x)\,,\,\tilde{J}^3_R(0,\y)]+ g^2[\tilde{J}^3_L(0,\x)\,,\,\tilde{J}^3_L(0,\y)]\\
&=&\frac{g^2}{2L^2}\sum_{N,M=-\infty}^{\infty}\Big([C^3_N\, , \, C^3_M]  e^{ik_N \x}e^{ik_M \y}
+[D^3_N\, , \, D^3_M]  e^{-ik_N \x}e^{-ik_M \y}\Big)\\
&=&\frac{g^2}{2L^2}\sum_{N=-\infty}^{\infty}\big(Ne^{ik_N( \x- \y)}+Ne^{-ik_N( \x- \y)}\big)=0
\eea

These relations imply that the Lagrange multipliers generate zero norm states
when applied to physical states. This is consistent with the expectation that
modes of the Lagrange multipliers can be found in zero norm physical states,
in analogy with the Gupta-Bleuler quantization of QED, where zero norm
combinations of the unphysical scalar and longitudinal photons are present
in the physical subspace.

\subsection{Quantization of the Bose field}

Our quantization of the Fermi Field follows standard methods.  The quantization 
of the gauge field involves more complex issues and is more delicate. In the case 
of the Schwinger modes the Bose field had to be quantized in indefinite metric~\cite{eliana}.  
For the free case that will also work here.  
Let us consider the part of the unperturbed Hamiltonian which involves the non-zero 
unphysical modes  of the gauge field:
\bean
H_G&=&\sum_{n=\half}^{\infty}\bigg\{ {b^+ _n}^\dag b^+ _n +{b^- _n}^\dag b^- _n
+i k_n {a^+_n}^\dag b^+ _n -i k_n { b^- _n}^\dag a^-_n
+i k_n {a^-_n}^\dag b^- _n -i k_n { b^+ _n}^\dag a^+_n \bigg\} \nn \\
&&+\sum_{N=1}^{\infty}\bigg\{ {b^3 _N}^\dag b^3 _N
+i k_N {a^3_N}^\dag b^3 _N -i k_N { b^3 _N}^\dag
 a^3_N \bigg\} \label{HG}
\; .
\eean
We are naturally led to a Fock representation with a vacuum state defined as the state  
$|0\rangle$  such that  $a^{3}_N  |0\rangle = b^{3}_N  |0\rangle =0 $
and $a^{\pm}_n  |0\rangle = b^{\pm}_n  |0\rangle =0 $
for $n, N >0$ . To implement that idea, following \cite{eliana}, we define, for $n, N >0$:
\bean
A_N ^3 &\equiv & \frac{1}{\sqrt{2L}}\,a^3 _N +i\sqrt {\frac{L}{2}}\,
b^3 _N  \\
A_{-N} ^3 &\equiv & \frac{1}{\sqrt{2L}}\,a^3 _N-i\sqrt {\frac{L}{2}}\,
b^3 _N  \\
A_n ^\pm &\equiv & \frac{1}{\sqrt{2L}}\,a^\pm _n +i\sqrt {\frac{L}{2}}\,
b^\pm _n  \label{Fock+n} \\
A_{-n} ^\pm &\equiv & \frac{1}{\sqrt{2L}}\,a^\mp _n-i\sqrt {\frac{L}{2}}\,
b^\mp _n  \label{Fock-n}\;
\eean
so that
\bea
a_N ^3 =\sqrt{\frac{L}{2}}\left(A^3 _N +A^3 _{-N}\right)\quad &,& \qquad
b_N ^3 = \frac{A^3 _N -A^3 _{-N}}{i\sqrt{2L}}\\
a_n ^\pm =\sqrt{\frac{L}{2}}\left(A^\pm _n +A^\mp _{-n}\right)\quad &,& \qquad
b_n ^\pm = \frac{A^\pm _n -A^\mp _{-n}}{i\sqrt{2L}}
\eea
The commutation relations
\bea
&&[A^3 _M, (A^3 _N)^\dag] =\delta_{MN} \quad , \qquad
[A^3 _{-M}, (A^3 _{-N})^\dag] =-\delta_{MN}\\
&& [A_m ^\pm , (A_ n ^\pm)^\dag]= \delta_{mn} \quad , \qquad
[A^\pm _{-m}, (A^\pm _{-n})^\dag] =-\delta_{mn} 
\eea 
can be represented in a Fock space endowed with an indefinite metric 
where the daggered operators are creation operators and the undaggered 
operators are destruction operators. As
a consequence of the unphysical nature of the degrees of freedom
we are considering, the presence of an indefinite metric is not
surprising and we know that it can be dealt with consistently
provided that its restriction to the physical subspace is positive
semidefinite. Note that (\ref{HG}) is not diagonal in this
representation, nor can it be diagonalized. The vacuum and the
states created out of it by repeated action of the operators
$(b^3_N)^\dag$ and $(b^\pm_n)^\dag$ provide an incomplete set of
eigenstates. This anomalous situation  is related to the fact that
the metric is not positive definite. A similar situation occurs in
the Schwinger model where it can, nonetheless,  be shown that  the
\emph{complete} set of one-particle eigenstates of the full
Hamiltonian, as given by the known solution of the model, can be
obtained perturbatively starting from the
\emph{incomplete }set of unperturbed one-particle eigenstates~\cite{eliana2}. 

The above quantization of the unphysical non-zero modes of the gauge
field is required for  the non-interacting gauge theory, where the subsidiary
conditions can be expressed as $b_N^3 |phys\rangle=0$ , $b_n^\pm |phys\rangle=0$ (as can easily be
seen by setting $g=0$ in the expressions for the Lagrange multipliers).
The physical subspace can be defined  by the 3 independent conditions 
\[ \left(A_N^3\! -\!A_{-N}^3\right)\!|phys\rangle \!=\!\left(A_n^\pm\! -\!A_{-n}^\pm \right)\!\!|phys\rangle\!=\!0\,,\]
expressed in terms of annihilation operators.
It is then possible to follow the  Gupta-Bleuler procedure and show that the physical subspace has a positive semi-definite metric, with zero-norm states being the ones containing ghost-like
modes, and  \\ $\langle phys| H_G| phys\rangle =0$,  so that unphysical modes do not contribute to the energy spectrum. 

As is characteristic of two-dimensional pure Yang-Mills theories in light-cone gauge, the Hamiltonian (\ref{HG}) has no interaction terms and  coincides with that of free gauge bosons. The interaction is carried by the
Lagrange multipliers and has the effect of modifying the subsidiary condition and the physical subspace.
We expect more restrictive conditions as a consequence of the g-dependent terms in $\lambda$.
The colour components of $\lambda$ do not commute with one another and the subsidiary conditions
are not independent. As a matter of fact, the Lagrange multipliers satisfy the same algebra as in the previously considered case
where fermions  are present. The conditions that need to be imposed are $\widetilde{\lambda}^3_0|phys\rangle=0$
and $\widetilde{\lambda}^3_N |phys\rangle=0$, for $N>0$ where
\bea
\widetilde{\lambda}^3_0 &=&
\frac{ig}{\sqrt{2}L}\,\sum_{m=\half}^{\infty}\bigg( (a^-_m)^\dag b^-_{m}+(b^+_m)^\dag a^+_{m}
-(a^+_m)^\dag b^+_{m}-(b^-_m)^\dag a^-_{m}\bigg)\\
\widetilde{\lambda}^3_N &=&\frac{i}{\sqrt{L}}\, k_N b_N^3
+\frac{ig}{\sqrt{2}L}\; \sum_{m=\half}^{N-\half}\left(b_m^- a^+_{N-m}-b_m^+ a^-_{N-m}\right)\nn \\
&+&\!\frac{ig}{\sqrt{2}L}\, \sum_{m=\half}^{\infty}\! \bigg( (a^-_m)^\dag b^-_{m+N}+(b^+_m)^\dag a^+_{m+N}
-(a^+_m)^\dag b^+_{m+N}-(b^-_m)^\dag a^-_{m+N}\bigg) \; .
\eea
We can  see that the eigenstates of (\ref{HG}) generated by the action of $ b_n^\pm$
are no longer physical states as in the free case. We need to require the more restrictive condition that no modes of $A^\pm$ be present in the physical subspace. Only physical states with modes of $A^3$  are now
zero-norm states and again one has  $\langle phys| H_G| phys\rangle =0.$ 

This indefinite metric representation of the gauge field, suggested by the free nature of the Hamiltonian
associated with it,  turns out to be unsuitable for the quantization of the full non-abelian gauge theory, on account
of its residual gauge invariance. To see this let us consider the operator
\be \label{U}
U(x)=e^{iN\pl (t+\x)\tau^3}
\ee
It satisfies the condition $U(t,-L)=U(t, L)$ and $A_- ^\prime =UA_-U^\dag +\frac{i}{g}\partial_-U U^\dag=0$.\\
It leaves the gauge-fixing condition invariant and it preserves
the boundary conditions. It is, therefore, a residual gauge
symmetry of the classical theory. Let us see what happens when we try to implement 
this symmetry in the quantized theory in the representation space where the gauge fields 
are quantized in indefinite metric.  To calculate the action of $u$ on $A$ we use
  $[\tau^+,\tau^-]=\tau^3$ and $[\tau^3, \tau^\pm]=\pm \tau^\pm$ to get:
\[
A ^\prime =UAU^\dag +\frac{i}{g}\partial_+U U^\dag = e^{iN\pl (t+\x)}A^- \tau^+ + e^{-iN\pl (t+\x)}A^+ \tau^-
-\frac{N\pi \sq}{gL}\tau^3
\]
or
\[
{A^-} ^\prime =e^{iN\pl (t+\x)}A^- \quad, \qquad {A^+} ^\prime =e^{-iN\pl (t+\x)}A^+ \quad, \qquad
{A^3} ^\prime=A^3-\frac{N\pi \sq}{gL}
\]
and from $F^\prime=UFU^\dag $
we have
\[
{F^-} ^\prime =e^{iN\pl (t+\x)}F^- \quad, \qquad {F^+} ^\prime =e^{-iN\pl (t+\x)}F^+ \quad, \qquad
{F^3} ^\prime=F^3 \;.
\]
Let us concentrate on the transformation properties of the + and $-$ colour components.  A quantum operator $T^N$ representing this symmetry in the space of states must be such that the quantum fields represented at
$t=0$ as in (\ref{Apm}) and (\ref{Fpm}) have the following transformation properties:
\bea
T^NA^- (0,\x) (T^N)^{\dag}&=&\frac{1}{\sqrt{L}} \,\sum_{n=-\infty}^{\infty}a_n^- e^{-in\pl\x}e^{iN\pl\x}=
\frac{1}{\sqrt{L}} \,\sum_{n=-\infty}^{\infty}a_{n+N}^- e^{-in\pl\x}\\
T^NA^+ (0,\x) (T^N)^{\dag}&=&\frac{1}{\sqrt{L}} \,\sum_{n=-\infty}^{\infty}a_n^+ e^{-in\pl\x}e^{-iN\pl\x}=
\frac{1}{\sqrt{L}} \,\sum_{n=-\infty}^{\infty}a_{n-N}^+ e^{-in\pl\x}\\
T^NF^-  (0,\x) (T^N)^{\dag}&=&\nor \,\sum_{n=-\infty}^{\infty} b^- _n e^{-in\pl \x}e^{iN\pl\x}=
\nor \,\sum_{n=-\infty}^{\infty} b^- _{n+N} e^{-in\pl \x}\\
T^NF^+  (0,\x) (T^N)^{\dag}&=&\nor \,\sum_{n=-\infty}^{\infty} b^+ _n e^{-in\pl \x}e^{-iN\pl\x}=
\nor \,\sum_{n=-\infty}^{\infty} b^+ _{n-N} e^{-in\pl \x}\;.
\eea
We see that must have
\bea
T^N a_n^+ (T^N)^{\dag}=a^+_{n-N} \quad, \qquad T^N b_n^+ (T^N)^{\dag}=b^+_{n-N} \\
T^N a_n^- (T^N)^{\dag}=a^-_{n+N} \quad, \qquad T^N b_n^- (T^N)^{\dag}=b^-_{n+N}
\eea
or, in terms of the Fock creation and annihilation operators defined in (\ref{Fock+n}--\ref{Fock-n}):
\bean
&&T^N A_n^+ (T^N)^{\dag}=A^+_{n-N} \qquad  \hspace{23pt} {\textstyle {\rm for} \ n\geq  N+\frac{1}{2}}\\
&&T^N A_{n}^+ (T^N)^{\dag}=(A^+_{n-N})^\dag
\hspace{10pt}\qquad {\textstyle {\rm for} \ \half \leq n \leq  N-\frac{1}{2}}\label{half1}\\
&&T^N A_{-n}^+ (T^N)^{\dag}=A^+_{-n-N}\qquad  \quad{\textstyle {\rm for} \ n\geq \frac{1}{2}}\\
&&T^NA_{-n}^- (T^N)^{\dag}=A^-_{-n+N}\hspace{12pt}\qquad {\textstyle {\rm for} \ n\geq  N+\frac{1}{2}}\\
&&T^N A_{-n}^- (T^N)^{\dag}=(A^-_{N-n})^\dag  \hspace{4pt}\qquad {\textstyle {\rm for} \ \half \leq n \leq  N-\frac{1}{2}}\label{half2}\\
&&T^N A_n^- (T^N)^{\dag}=A^-_{n+N}\hspace{11pt}\qquad  \quad{\textstyle {\rm for} \ n\geq \frac{1}{2}}
\eean
We can see from (\ref{half1}) and (\ref{half2}) that $T$ must turn annihilation operators into 
creation operators.
This does not allow the vacuum to be invariant. The transformed vacuum $T|0\rangle$
must be such that $(A^+_{-\half})^\dag T|0\rangle=0$, a condition which cannot be satisfied 
by a state in the Fock space we are considering. The symmetry of the theory
under index-shifting at  a classical level suggests that in a Fock quantization the 
creation or annihilation nature of the Bose operators must be preserved under index-shifting 
(that need not be true for Fermi operators where a relation such as 
$(\delta_{-\half})^\dag T|0\rangle=0$ is easily satisfied). Interpreting $(A^\pm_{-n})^\dag$ as creation 
operators generating negative-norm states when acting on the vacuum appears to be 
inconsistent with  this symmetry
transformation. As we shall see, implementing this symmetry as a unitary operator in the Hilbert 
space will be necessary to obtain the non-trivial vacuum
structure which is characteristic of this theory when coupled to fermions.  One possibility is that 
the relation $(A^+_{-\half})^\dag T|0\rangle=0$ must be implemented weakly in the physical 
subspace so that the relation holds in the factor space that forms the physical Hilbert space.  We 
do not know whether or not that idea can be realized and we will not pursue it further in this paper.
Here we shall solve the problem by modifying the quantization of the fields $A^\pm$.

In the quantization of $A^\pm$ in the free gauge theory, as well as  of  $A^3$  in both the free and
the interacting case, the indefinite metric is necessary to express the subsidiary
condition in terms of annihilation operators and it allows to get rid of ghost-like modes, which are 
present in zero-norm physical states, by constructing a Hilbert space with positive definite metric 
as a quotient space.  But in the
quantization of the interacting $A^\pm$ the indefinite metric does not seem to play a crucial role. 
As a matter of fact, a standard definition of creation and annihilation operators with canonical commutators:
\[
 A_n ^\pm \equiv  \frac{1}{\sqrt{2L}}\,a^\pm _n +i\sqrt {\frac{L}{2}}\,
b^\pm _n \ ,   \qquad [A_m ^\pm , (A_ n ^\pm)^\dag]= \delta_{mn} \ ,
\]
for both positive and negative $n$, leads to the following expressions for $\widetilde{\lambda}_0^3 $ and
$\widetilde{\lambda}_N^3$:
\bea
\widetilde{\lambda}^3_0 &=&
\frac{g}{\sqrt{2}L}\sum_{m=-\infty}^{\infty}\bigg( (A_{m}^-)^\dag A_m^-
- (A_{m}^+)^\dag A_m^+ \bigg)\\
\widetilde{\lambda}^3_N &=&\frac{i}{\sqrt{L}}\, k_N b_N^3
+\frac{g}{\sqrt{2}L}\sum_{m=-\infty}^{\infty}\bigg( (A_{m-N}^-)^\dag A_m^-
- (A_{m-N}^+)^\dag A_m^+ \bigg)\; .
\eea
The subsidiary conditions can be satisfied by requiring that
\bea
&&(A_N^3-A_{-N}^3) |phys\rangle =0\ , \forall N>0 \\
&&A_n^\pm |phys\rangle=0
\eea
and (\ref{HG}) can be written as
\bea
H_G&=&\sum_{n=-\infty}^{\infty}\left\{ \left(k_n+\frac{1}{2L}\right) \bigg( (A_n^+)^\dag A_n^+ +(A_n^-)^\dag A_n^- \bigg)
-(A_n^+)^\dag (A_{-n}^-)^\dag -A_n^+ A_{-n}^- \right\} \\
&&+\sum_{N=1}^{\infty}\bigg\{ {b^3 _N}^\dag b^3 _N
+i k_N {a^3_N}^\dag b^3 _N -i k_N { b^3 _N}^\dag
 a^3_N \bigg\} \; .
\eea
The vacuum is the only physical state in the positive metric Fock representation of $A^+$ and $A^-$
and, although it is not an eigenstate of $H_G$,  we still have $\langle 0| H_G|0\rangle=0$.
The transformation $U$ can now be represented by an operator $T^N$ such that:
\bean
&&T^N A_n^+ (T^N)^{\dag}=A^+_{n-N}\quad, \quad \qquad
T^N (A_n^+)^\dag (T^N)^{\dag}=(A^+_{n-N})^\dag \label {TA+}\\
&&T^N A_n^- (T^N)^{\dag}=A^-_{n+N}\quad, \quad \qquad
T^N (A_n^+)^\dag (T^N)^{\dag}=(A^+_{n-N})^\dag \label{TA-}
\eean
for any positive or negative $n$.\\
One can check that (\ref{TA+}--\ref{TA-}) are satisfied for $N=1$ by
\newpage
\beal \label{operator}
\widetilde{T}=&\cdots &e^{\frac{\pi}{2}\left( A_n^ + (A^+_{n-1})^\dag - (A_n^+)^\dag  A^+_{n-1}\, +\,
A_{-n}^- (A^-_{-n+1})^\dag - (A_{-n}^-)^\dag A^-_{-n+1}\right)}\cdots \\
&\cdots& e^{\frac{\pi}{2}\left( A_{ 3/2}^+ (A^+_{1/2})^\dag - (A_{3/2}^+)^\dag  A^+_{1/2}\,+\,
A_{-3/2}^- (A^-_{-1/2})^\dag - (A_{-3/2}^-)^\dag A^-_{-1/2}\right)}\\
&& e^{\frac{\pi}{2}\left( A_{ 1/2}^+ (A^+_{-1/2})^\dag - (A_{1/2}^+)^\dag  A^+_{-1/2}\,+\,
A_{-1/2}^- (A^-_{1/2})^\dag - (A_{-1/2}^-)^\dag A^-_{1/2}\right)} \\
&& e^{\frac{\pi}{2}\left( A_{- 1/2}^+ (A^+_{-3/2})^\dag - (A_{-1/2}^+)^\dag  A^+_{-3/2}\,+\,
A_{1/2}^- (A^-_{3/2})^\dag - (A_{1/2}^-)^\dag A^-_{3/2}\right)} \cdots \\
&\cdots &e^{\frac{\pi}{2}\left( A_{-n}^ + (A^+_{-n-1})^\dag - (A_{-n}^+)^\dag  A^+_{-n-1}\, +\,
A_{n}^- (A^-_{n+1})^\dag - (A_{n}^-)^\dag A^-_{n+1}\right)}\cdots
\eeal

\subsection{A Hamiltonian Suitable for Perturbative Calculations}

Although unphysical, the modes of $A^+$ and $A^-$ interact with fermions and can no longer be
eliminated from the theory when coupling to fermions is considered.
The fact that the gauge-invariant vacuum state associated with a positive-definite Fock
representation of $A^+$ and $A^-$ is not an eigenstate of the unperturbed Hamiltonian
prevents us from performing a standard perturbative calculation.  On the other hand,
 as a result of the gauge-invariant renormalization of the fermion products, the term
$\, \frac{g^2}{2\pi}A^+ A^- \,$ has been introduced into the theory. Being quadratic in
the gauge field it has the well known form of a mass term, the ``mass'' being $m\equiv \frac{g}{\sqrt{\pi}}$. This suggests treating it as part of the unperturbed Hamiltonian,
in spite of its dependence on $g^2$, leaving only  the order-g terms $g (A^+  \widetilde{J}^-_R +
A^-  \widetilde{J}^+_R) \, $,
which couple the gauge fields to the fermion currents, in the perturbation
Hamiltonian.  By doing so we can diagonalize the ``unperturbed" hamiltonian related to the $+$ and $-$ gauge fields. We can write
\bea
H_G=\sum_{n=-\infty}^{\infty}(k_n+m)\left( (A^+_n)^\dag A^+_n + (A^-_n)^\dag A^-_n\right)
+\sum_{N=1}^{\infty}\bigg\{ {b^3 _N}^\dag b^3 _N
+i k_N {a^3_N}^\dag b^3 _N -i k_N { b^3 _N}^\dag
 a^3_N  \bigg\}
\eea
where now the Fock operators $A_n^\pm$ are defined as
\[
A_n ^\pm  \equiv  \sqrt{\frac{m}{2}}\,a^\pm _n +\frac{i}{\sqrt{2m}}\,b^\pm _n
\]
for both positive and negative $n$, with  $m=\frac{g}{\sqrt{\pi}}$.\\
Inverting these relations we obtain
\bea
a_n ^\pm =\frac{A^\pm _n +(A^\mp _{-n})^\dag}{\sqrt{2m}}\quad &,& \qquad
b_n ^\pm =\sqrt{\frac{m}{2}}\, \frac{A^\pm _n -(A^\mp _{-n})^\dag}{i}\;.
\eea
From (\ref{commab}) we see that these operators satisfy the Fock algebra
\[
[A_m ^\pm , {A_ n ^\pm}^\dag]= \delta_{mn}
\]
all the other commutators vanishing.\\
Analogously, we quantize the zero mode of $A^3 $ by defining
\[
A_0^3 \equiv  \sqrt{\frac{m}{2}}\,a^3 _0 +\frac{i}{\sqrt{2m}}\,b^3 _0
\]
so that we can write
\[
\half \left(b_0 ^3\right)^2 +\frac{g^2}{2\pi}\left(a_0 ^3\right)^2=
m{A_0^3}^\dag A_0^3 \;.
\]
The Hamiltonian can now be written as
\[
\hat{H}= H_0+H_I
\]
where
\bea
H_0=H_F^0 +
\sum_{N=1}^{\infty}\bigg\{ {b^3 _N}^\dag b^3 _N
+i k_N {a^3_N}^\dag b^3 _N -i k_N { b^3 _N}^\dag
 a^3_N  \bigg\} +m (A_0^3)^\dag A_0^3\\
+\sum_{n=-\infty}^{\infty}
\Big(k_n+m\Big) \bigg( (A_n^+)^\dag A_n^+ + (A_n^- )^\dag A_n^- \bigg)
\eea
and
\bea
H_I=
\sum_{N=1}^{\infty}\bigg\{
g\sqrt{\frac{2}{L}}\left(C^3 _N a^3 _N +{C^3 _N}^\dag {a^3 _N}^\dag\right)
 +\frac{g^2}{\pi}{a^3_N}^\dag a^3_N\bigg\} \Bigg\} +\frac{g}{\sqrt{mL}}C_0^3 \Big((A_0^3)^\dag + A_0^3 \Big)\\
+ \frac{g}{\sqrt{mL}}\, \Bigg\{ \sum_{n=-\infty}^{\infty}\bigg( A_n^- C_n^+ +A_n^+ C_n^- +(A_n^-)^\dag (C_n^+)^\dag +(A_n^+)^\dag (C_n^-)^\dag \bigg)
\eea
$H_F^0$ is the free fermion Hamiltonian (\ref{HF}-- \ref{HFbos}).

\section{Perturbative Calculations}

\subsection{The Vacuum}

Let
\bea
|\Omega\rangle &=& |0\rangle + |\Omega^{(1)}\rangle + |\Omega^{(2)}\rangle +\dots\\
E&=&E_0 + E_1 + E_2 +\dots
\eea
We shall determine the corrections to the vacuum state, $ |\Omega^{(1)}\rangle$ and $ |\Omega^{(2)}\rangle$, and to the
vacuum energy, $E_1$ and $E_2$, by requiring that
\bean
&&H_0 |0\rangle = E_0  |0\rangle\\
&&H_0  |\Omega^{(1)}\rangle + H_1  |0\rangle = E_0  |\Omega^{(1)}\rangle + E_1  |0\rangle \label{first} \\
&&H_0  |\Omega^{(2)}\rangle +  H_1 |\Omega^{(1)}\rangle = E_0  |\Omega^{(2)}\rangle +  E_1 |\Omega^{(1)}\rangle +E_2  |0\rangle \;.\label{second}
\eean
$H_0$ is normal-ordered in such a way that
\[
H_0  |0\rangle =  0 \
\quad {\rm and}\quad  E_0=0
\]
while higher order corrections to the energy are not necessarily zero.
 We have
\bea
H_I |0\rangle=
\sum_{N=1}^{\infty}
g\sqrt{\frac{2}{L}}(C^3 _N)^\dag {a^3 _N}^\dag|0\rangle  
 +\frac{g}{\sqrt{mL}}\sum_{n=-\infty}^{\infty}\bigg((A_n^-)^\dag  (C_n^+)^\dag |0\rangle +(A_n^+)^\dag (C_n^-)^\dag |0\rangle \bigg)  \,.
\eea
One can immediately see that
\[
E_1=\langle 0|H_I |0\rangle =0\,.
\]
As we have seen,  $\ [H_F\, , \, C_n^+]=-k_n C_n^+$,  so that we have
\[
[H_0 \, , \, C_n^+]=-k_n C_n^+ \quad {\rm and } \quad [H_0 \, , \, (C_n^+)^\dag]=k_n (C_n^+)^\dag \, .
\]
Analogously one can prove that
\[
[H_0 \, , \, C_n^-]=-k_n C_n^- \quad {\rm and } \quad [H_0 \, , \, (C_n^-)^\dag]=k_n (C_n^-)^\dag \, .
\]
These relations, together with $H_0 |0\rangle=0$ , tell us that
\[
H_0 (C_n^\pm )^\dag |0\rangle = k_n (C_n^\pm )^\dag |0\rangle\, ,
\]
while from the expressions of $C_n^\pm$ (\ref{Cn+}--\ref{Cn-}) we see that
\[
C_n^\pm |0\rangle = (C_{-n}^\mp)^\dag |0\rangle = 0 \,.
\]
We also have $C_N^3 |0\rangle = (C_{-N}^3)^\dag \0 = 0$ and, from (\ref{HFbos}),
\[
H_0 (C_N^3)^\dag  |0\rangle = k_N (C_N^3)^\dag  |0\rangle \,.
\]
It is now easy to verify that the state
\bea
|\Omega^{(1)}\rangle
&=&
-\frac{g \sq}{\sqrt{L}}\sum_{N=1}^{\infty}\left(\frac{{a_N^3}^\dag}{2k_N}+\frac{i{b_N^3}^\dag}{(2k_N)^2}
\right){C_N^3}^\dag \0 \\
&&-\frac{g}{\sqrt{mL}}\sum_{n=\half}^{\infty}\frac{1}{2k_n+m}\bigg( (A_n^-)^\dag (C_n^+)^\dag |0\rangle +(A_n^+)^\dag (C_n^-)^\dag |0\rangle \bigg)
\eea
satisfies (\ref{first}) with $E_0=E_1=0$ .\\
In order to evaluate $E_2$ and $|\Omega^{(2)}\rangle$ we need $H_I |\Omega^{(1)}\rangle$. Using (\ref{commc3}), (\ref{commcpcm}) we get
\bea
H_I |\Omega^{(1)}\rangle \!\!\!&=&\!\!\!
\frac{2g^2}{L} \sum_{N=1}^{\infty}\frac{N}{(2k_N)^2} |0\rangle
-\frac{g^2}{mL}\sum_{n=\half}^{\infty}\frac{2n}{2k_n+m}|0\rangle  \\
\! \!\!&-&\!\!\!\frac{2g^2}{L} \sum_{M,N=1}^{\infty}{a_M^3 }^\dag {C_M^3}^\dag
\left(\frac{{a_N^3}^\dag}{2k_N}+\frac{i{b_N^3}^\dag}{(2k_N)^2}
\right){C_N^3}^\dag \0\\
\!\!\!&-&\!\!\! \frac{g^2 \sq}{\sqrt{m}L}\sum_{M=1}^{\infty}\sum_{n=\half}^{\infty}
\frac{{a_M^3}^\dag (C_M^3)^\dag \bigg( (A_n^-)^\dag  (C_n^+)^\dag +(A_n^+)^\dag (C_n^-)^\dag  \bigg)}{2k_n+m}|0\rangle\\
\!\!\!&-&\!\!\!  \frac{g^2 \sq}{\sqrt{m}L} \sum_{m=-\infty}^{\infty}\sum_{N=1}^{\infty}\bigg((A_m^-)^\dag   (C_m^+)^\dag +(A_m^+)^\dag (C_m^-)^\dag \bigg) \left(\frac{{a_N^3}^\dag}{2k_N}+\frac{i{b_N^3}^\dag}{(2k_N)^2}
\right){C_N^3}^\dag \0 \\
\!\!\!&-&\!\!\! \frac{g^2}{mL}\sum_{n=\half}^{\infty}\frac{ (A_0^3)^\dag
 C_0^3 \bigg( (A_n^-)^\dag  (C_n^+)^\dag +(A_n^+)^\dag (C_n^-)^\dag  \bigg)}{2k_n+m}|0\rangle\\
\!\!\!&-&\!\!\! \!\! \frac{g^2}{mL} \sum_{m=-\infty}^{\infty}\sum_{n=\half}^{\infty}\frac{\bigg((A_m^-)^\dag   (C_m^+)^\dag +(A_m^+)^\dag (C_m^-)^\dag  \bigg) \bigg((A_n^-)^\dag   (C_n^+)^\dag +(A_n^+)^\dag (C_n^-)^\dag  \bigg)}{2k_n+m}\0
 \;.
\eea
 We therefore have
\[
E_2=
\frac{2g^2}{L} \sum_{N=1}^{\infty}\frac{N}{(2k_N)^2}
-\frac{g^2}{mL}\sum_{n=\half}^{\infty}\frac{2n}{2k_n+m}
\]
a diverging quantity that has to be subtracted from the Hamiltonian. \\
It is not hard  to verify  that (\ref{second}) is satisfied if the state $|\Omega^{(2)}\rangle$ is given by

\baselineskip18pt
\bea \label{2}
|\Omega^{(2)}\rangle  \!\!\!&=&\!\!\!
\frac{g^2}{L}\, \sum_{M=1}^{\infty}\sum_{N=1}^{\infty}
\left(\frac{{a_M^3}^\dag}{2k_M}+\frac{i{b_M^3}^\dag}{(2k_M)^2}
\right)\left(\frac{{a_N^3}^\dag}{2k_N}+\frac{i{b_N^3}^\dag}{(2k_N)^2}
\right){C_M^3}^\dag {C_N^3}^\dag \0 \\ \\
&\;+&\!\!\!\! \frac{g^2 \sq}{\sqrt{m}L}\sum_{M=1}^{\infty}\sum_{n=\half}^{\infty}\Bigg(\frac{{a_M^3}^\dag}{2k_{M}+2k_n +m}\\
&&\hspace{70pt}+\frac{i{b_M^3}^\dag}{(2k_{n}+2k_M +m)^2} \Bigg){C_M^3}^\dag \, \frac{ (A_n^-)^\dag  (C_n^+)^\dag +(A_n^+)^\dag (C_n^-)^\dag }{2k_n+m}\0 \\ \\
 &\;+&\!\!\!\! \frac{g^2 \sq}{\sqrt{m}L}\sum_{N=1}^{\infty}\sum_{p=-\infty}^{\infty}
\Bigg\{ \frac{1}{2k_{p}+2k_N +m} \left(\frac{{a_N^3}^\dag}{2k_N}+\frac{i{b_N^3}^\dag}{(2k_N)^2}
\right) \\
&&\hspace{65pt}+ \frac{i{b_N^3}^\dag}{2k_N (2k_{p}+2k_N +m)^2}\Bigg\}
\bigg((A_p^-)^\dag   (C_p^+)^\dag +(A_p^+)^\dag (C_p^-)^\dag  \bigg){C_N^3}^\dag \0\\ \\
&\;+&\!\!\!\!\frac{g^2}{mL}\sum_{n=\half}^{\infty} \frac{ (A_0^3)^\dag
 \bigg({- (A_n^-)^\dag  (C_n^+)^\dag +(A_n^+)^\dag (C_n^-)^\dag }\bigg)}
{2(k_{n}+m) (2k_n+m)}|0\rangle\\ \\
&\,+&\!\!\!\! \!\frac{g^2}{mL} \sum_{p=\half}^{\infty}\sum_{n=\half}^{\infty} \left[\frac{(A_p^+)^\dag (A_n^+)^\dag(C_p^-)^\dag   (C_n^-)^\dag  }{2(2k_p +m)(2k_n+m)}\0
+\frac{(A_p^-)^\dag (A_n^-)^\dag(C_p^+)^\dag   (C_n^+)^\dag  }{2(2k_p +m)(2k_n+m)}\0\right]\\ \\
&+&\!\!\!\! \! \!\frac{g^2}{mL}\sum_{p=-\infty}^{\infty}\sum_{n=\half}^{\infty} \left[ \frac{(A_p^-)^\dag (A_n^+)^\dag(C_p^+)^\dag   (C_n^-)^\dag  }{2(k_{n}+k_p +m)(2k_n+m)}\0
+\frac{(A_p^+)^\dag (A_n^-)^\dag(C_p^-)^\dag   (C_n^+)^\dag  }{2(k_{n}+k_p +m)(2k_n+m)}\0\right]
.
\eea

\subsubsection{The subsidiary condition}

Let us verify that the state $|\Omega \rangle = \0 +| \Omega^{(1)}\rangle +|\Omega^{(2)}\rangle $  satisfies the subsidiary condition.
It is easy to see that
\[ \lambda_0^3 |\Omega \rangle =0
\]
where
 \[
\lambda_0^3 =\frac{g}{\sq L}\left( C_0^3+ D_0^3 \right)+\frac{g}{\sq L}\sum_{m=-\infty}^{\infty}\bigg( (A_{m}^-)^\dag A_m^-
- (A_{m}^+)^\dag A_m^+ \bigg) \,.
\]
As a matter of fact we have
\bea
&&\lambda_0^3 \0 =  0 \\
&&\lambda_0^3 \1 = -\frac{g^2}{L \sqrt{2mL}} \, \Bigg\{ \sum_{n=\half}^{\infty}\frac{C_0^3}{2k_n+m}\bigg( (A_n^-)^\dag (C_n^+)^\dag |0\rangle +(A_n^+)^\dag (C_n^-)^\dag |0\rangle \bigg)\\
&& \hspace{60pt}+\sum_{n=\half}^{\infty}\frac{1}{2k_n+m}\bigg( (A_n^-)^\dag (C_n^+)^\dag |0\rangle -(A_n^+)^\dag (C_n^-)^\dag |0\rangle \bigg) \Bigg\}
\eea
Using  the relation $ [C_0^3, (C_n^\pm)^\dag ] =\mp (C_{n}^\pm)^\dag$ it is easy
to see that $\lambda_0^3 \1 =0$ and that the same holds for every term in $\2$
(p. \pageref{2}). Each term in the perturbative expansion of the physical vacuum
$  |\Omega \rangle $ is an eigenstate of  the conserved charge $\lambda_0^3  $ with
eigenvalue $0$. As we have seen, in order to be a physical state, the vacuum must also be annihilated by the positive frequency components of $\lambda^3$. This means that it must satisfy $ \lambda^3_N  |\Omega \rangle =0 $ where
\bea
\lambda^3_N &=&\frac{ik_N b_N^3}{\sqrt{L}}+ \frac{g}{\sq L}\left( (C_N^3)^\dag + D_N^3 \right)\\
&&+\frac{g}{\sq L}\sum_{m=-\infty}^{\infty}\bigg( (A_{m-N}^-)^\dag A_m^-
- (A_{m-N}^+)^\dag A_m^+ \bigg)\; .
\eea
Clearly this condition cannot be satisfied term by term in the expansion of
$  |\Omega \rangle $,  as is the case for $\lambda_0^3$. The action of   $\lambda_N^3$
mixes up the perturbative orders and the condition cannot be satisfied exactly
by our perturbative evaluation of the vacuum. We can only check that it holds for
the two lowest orders in the expansion of  $\lambda_N^3 |\Omega \rangle$.
We have
\bea
&&\lambda_N^3 \0 =\frac{g}{\sq L} (C_N^3)^\dag \0 \\
&&\lambda_N^3 \1 = -\frac{g}{\sq L}
(C_N^3)^\dag \0 -\frac{g^2}{L \sqrt{L}}\sum_{M=1}^\infty \left(\frac{{a_M^3}^\dag}{2k_M}+\frac{i{b_M^3}^\dag}{(2k_M)^2}
\right) (C_N^3)^\dag (C_M^3)^\dag \0 \\
&&\hspace{60pt}-\frac{g^2}{L \sqrt{2mL}}\sum_{n=\half}^\infty \frac{(C_N^3)^\dag}{2k_n +m}
 \left( (C_n^- )^\dag (A_n^+ )^\dag + (C_n^+ )^\dag (A_n^- )^\dag \right) \0 \\
&&\hspace{60pt}-\frac{g^2}{L \sqrt{2mL}}\sum_{n=\half}^\infty \frac{1}{2k_{n}+m}\left( (C_{n}^+)^\dag (A_{n-N}^-)^\dag - (C_{n}^-)^\dag (A_{n-N}^+)^\dag \right)\0  \,.
  \eea
Disregarding higher order terms we can write
\bea
\lambda_N^3 \1 \simeq -\frac{g}{\sq L} (C_N^3)^\dag \0
-\frac{g^2}{L \sqrt{L}}\sum_{M=1}^\infty \left(\frac{{a_M^3}^\dag}{2k_M}+\frac{i{b_M^3}^\dag}{(2k_M)^2}
\right) (C_N^3)^\dag (C_M^3)^\dag \0 \\
-\frac{g^2}{L\sqrt{2mL}}\Bigg\{
 \sum_{n=\half}^\infty \frac{(C_N^3)^\dag}{2k_n }
 \left( (C_n^- )^\dag (A_n^+ )^\dag + (C_n^+ )^\dag (A_n^- )^\dag \right) \0 \hspace{10pt}\\
\hspace{65pt}+\sum_{n=\half}^\infty \frac{1}{2k_{n}}\left( (C_{n}^+)^\dag (A_{n-N}^-)^\dag - (C_{n}^-)^\dag (A_{n-N}^+)^\dag \right)\0 \Bigg\}
\eea
and keeping only the lowest order terms in $\lambda_N^3 \2 $ we get
\bea
\lambda_N^3 \2 &\simeq &
\frac{g^2}{L \sqrt{L}}\sum_{M=1}^\infty \left(\frac{{a_M^3}^\dag}{2k_M}+\frac{i{b_M^3}^\dag}{(2k_M)^2}
\right) (C_N^3)^\dag (C_M^3)^\dag \0 \\
&&+\frac{g^2}{L\sqrt{2mL}}\Bigg\{
 \sum_{n=\half}^\infty \frac{k_N (C_N^3)^\dag \bigg( (A_n^+ )^\dag(C_n^- )^\dag +
(A_n^- )^\dag(C_n^+ )^\dag \bigg)}{2k_n k_{n+N}} \0 \\
&&\hspace{50pt}+ \sum_{n=-\infty}^\infty \frac{ \bigg( (A_n^+ )^\dag(C_n^- )^\dag +
(A_n^- )^\dag(C_n^+ )^\dag \bigg)(C_N^3)^\dag}{2 k_{n+N}} \0 \Bigg\} \,.
\eea
Using the relation $ [(C_n^\pm)^\dag  , (C_N^3)^\dag] =\pm (C_{n+N}^\pm)^\dag$ we get
\bea
 \lambda_N^3 \2 &\simeq &
\frac{g^2}{L \sqrt{L}}\sum_{M=1}^\infty \left(\frac{{a_M^3}^\dag}{2k_M}+\frac{i{b_M^3}^\dag}{(2k_M)^2}
\right) (C_N^3)^\dag (C_M^3)^\dag \0 \\
&&+\frac{g^2}{L\sqrt{2mL}}\Bigg\{
 \sum_{n=\half}^\infty \frac{1}{2k_n } (C_N^3)^\dag \bigg( (A_n^+ )^\dag(C_n^- )^\dag +
(A_n^- )^\dag(C_n^+ )^\dag \bigg) \0 \\
&&\hspace{50pt}+ \sum_{n=-N+\half}^\infty \frac{1}{2 k_{n+N}} \bigg( -(A_n^+ )^\dag(C_{n+N}^- )^\dag +
(A_n^- )^\dag(C_{n+N}^+ )^\dag \bigg) \0 \Bigg\}
\eea
and one can immediately see that
\[\lambda_N^3 \Big(\0 + \1 +\2 \Big) \simeq 0 \; . \]
Note that this condition would not be satisfied even for the two lowest orders
if $A^3$ had been quantized with a positive definite metric,
while no such inconsistency appears as a consequence of our quantization
of $A^+$ and $A^-$.

\subsubsection{The degenerate vacua}

We have seen that under the residual gauge transformation
\be
U(x)=e^{iN\pl (t+\x)\tau^3}
\ee
the gauge fields transform according to
\bea
&&{A^-} ^\prime =e^{iN\pl (t+\x)}A^- \quad, \qquad {A^+} ^\prime =e^{-iN\pl (t+\x)}A^+ \quad, \qquad
{A^3} ^\prime=A^3-\frac{N\pi \sq}{gL}\\
&&{F^-} ^\prime =e^{iN\pl (t+\x)}F^- \quad, \qquad {F^+} ^\prime =e^{-iN\pl (t+\x)}F^+ \quad, \qquad
{F^3} ^\prime=F^3\;.
\eea
As a consequence, under the action of the operator $T^N$ representing this transformation, the
Fourier modes into which the fields are decomposed at $t=0$ transform as
\bean
&&T^N a_n^+ (T^N)^{\dag}=a^+_{n-N} \quad, \qquad \hspace{34pt} T^N b_n^+ (T^N)^{\dag}=b^+_{n-N} \label{ta+t}\\
&&T^N a_n^- (T^N)^{\dag}=a^-_{n+N} \quad, \qquad  \hspace{34pt}  T^N b_n^- (T^N)^{\dag}=b^-_{n+N}\label{ta-t}\\
&&T^N a_M^3 (T^N)^{\dag}=a^3_{M} \quad {\rm for}\ M\neq 0 \\
&&T^N a_0^3 (T^N)^{\dag}=a^3_{0}-\frac{N\pi \sq}{g\sqrt{L}}\label{ta0t}\\
&&T^N b_M^3 (T^N)^{\dag}=b^3_{M}\;.
\eean
From (\ref{ta+t}) and (\ref{ta-t}) we also get
\bean
&&T^N A_n^+ (T^N)^{\dag}=A^+_{n-N}\\
&&T^N A_n^- (T^N)^{\dag}=A^-_{n+N}
\eean
Let us consider the transformation of the Fermi field. It is easy to see that
\[
\Psi^\prime = U\Psi U^\dag = e^{iN\pl (t+\x)}\psi \tau^+ + e^{-iN\pl (t+\x)}\psi^\dag \tau^- +\phi \tau^3
\;.\]
Therefore, at $t=0$,
\[
\psi^\prime (0,\x) =e^{iN\pl \x}\psi (0,\x) \quad, \qquad \phi^\prime(0,\x) = \phi (0,\x)
\]
From the bosonized form of $\psi_{R/L}$ (eqs. \ref{psiRbos}, \ref{psiLbos}), using relations (\ref{sigmasigma+R}---\ref{1.28}) and the identity
\[
e^A B=Be^A e^c  \quad {\rm if}\  [A, B]=cB , \quad {\rm where}\ c \ {\rm is \ a}\ {\rm c-number}
\; ,\]
one can easily prove that
\[
\sigma_L^\dag \sigma_R e^{i\pi (C_0^3 + D_0^3)}\psi (0,\x) \left(\sigma_L^\dag \sigma_R e^{i\pi (C_0^3 + D_0^3)}\right)^\dag =e^{iN\pl \x}\psi (0,\x) \;.
\]
We also have
\[
e^{-\frac{i\pi\sq}{g\sqrt{L}}b_0^3}a_0^3\, e^{\, \frac{i\pi\sq}{g\sqrt{L}}b_0^3}
={\textstyle a_0^3-\frac{\pi\sq}{g\sqrt{L}} }\;.
\]
The operator $T$ representing the residual gauge transformation can therefore
be expressed as
\[
T=\sigma_L^\dag \sigma_R e^{i\pi (C_0^3 + D_0^3)}e^{-\frac{i\pi\sq}{g\sqrt{L}}b_0^3}\,\widetilde{T}\;.
\]
\hspace{-7pt}
where $\widetilde{T}$ is the operator transforming $A^+$ and $A^-$ under the residual gauge
transformation.
In order to determine the transformation properties of $C^\pm_n$ let us consider
$\psi^\prime_R$.  We can write
\bea
\psi^\prime _R(0,\x)&=& {1 \over \sqrt {2L}} \sum_{n=\half}^\infty
\left(b_n e^{in\pl \x}e^{iN\pl \x} + d{_n ^\dag} e^{-in\pl \x}e^{iN\pl \x} \right)\\
&=&{1 \over \sqrt {2L}}\Bigg( \sum_{n={N+\half}}^\infty
b_{n-N} e^{in\pl \x} + \sum_{n=-N+\half}^\infty d{_{n+N} ^\dag} e^{-in\pl \x}\Bigg) \;.
\eea
For $N>0$  $\psi^\prime _R$ can be written as
\[
\psi^\prime _R (0,\x)= {1 \over \sqrt {2L}}\left( \sum_{n={N+\half}}^\infty
b_{n-N} e^{in\pl \x} + \sum_{n=\half}^{N-\half} d{_{N-n} ^\dag} e^{in\pl \x}
+\sum_{n=\half}^\infty d{_{n+N} ^\dag} e^{-in\pl \x}\right)
\]
and for $N<0$
\[
\psi^\prime _R (0,\x)= {1 \over \sqrt {2L}}\left( \sum_{n=\half}^\infty
b_{n-N} e^{in\pl \x} + \sum_{n=\half}^{-N-\half} b_{-n-N}  e^{-in\pl \x}
+\sum_{n=-N+\half}^\infty d{_{n+N} ^\dag} e^{-in\pl \x}\right)\;.
\]
We can therefore see that we must have
\bea
T^N b_n (T^N)^{\dag}&=&b_{n-N}\quad \ {\rm for} \ N<n \\
T^N b_n (T^N)^{\dag}&=&d_{N-n}^\dag  \quad {\rm for} \ N>n \\
T^N d_n (T^N)^{\dag}&=&d_{n+N}^\dag  \quad \ {\rm for} \ N>-n \\
T^N d_n (T^N)^{\dag}&=&b_{-N-n}^\dag  \quad {\rm for} \ N<-n
\;.
\eea
We can now determine the transformation properties of $C_n^\pm$. \\
For $n>N>0$ we have
\bea
T^N C_n^- (T^N)^{\dag}&=&\sum_{m=\half}^{\infty}d_{m+N}^\dag r_{n+m}
-\sum_{M=0}^\infty r_M^\dag b_{M+n-N}\\
&&-\sum_{m=N+\half}^n r_{n-m}b_{m-N}-\sum_{m=\half}^{N-\half}r_{n-m}
d_{N-m}^\dag\\
&=&\sum_{m=N+\half}^{\infty}d_{m}^\dag r_{n-N+m}
-\sum_{M=0}^\infty r_M^\dag b_{M+n-N}\\
&&-\sum_{m=\half}^{n-N} r_{n-N-m}b_{m}+\sum_{m=\half}^{N-\half}
d_{m}^\dag r_{n-N+m}\\
&=&C_{n-N}^-
\eea
and for $N>n>0$
\bea
T^N C_n^- (T^N)^{\dag}&=&\sum_{m=\half}^{\infty}d_{m+N}^\dag r_{n+m}
-\sum_{M=0}^{N-n-\half} r_M^\dag d_{N-n-M}^\dag\\
&&-\sum_{M=N-n+\half}^\infty r_{M}^\dag b_{M+n-N}-\sum_{m=\half}^{n}r_{n-m}
d_{N-m}^\dag\\
&=&\sum_{M=n+\half}^{\infty}d_{M-n+N}^\dag r_{M}
+\sum_{M=0}^{n-\half}d_{M-n+N}^\dag r_M \\
&&-\sum_{m=\half}^{N-n} r_{N-n-m}^\dag d_{m}^\dag -\sum_{m=\half}^{\infty}
r_{m+N-n}^\dag b_{m} \\
&=&(C_{N-n}^+ )^\dag \equiv C_{n-N}^-
\eea
Analogously it is possible to show that for any positive or negative $N$
\bean
T^N C_n^- (T^N)^{\dag}&=&C_{n-N}^-  \label{tc-t}\\
T^N C_n^+ (T^N)^{\dag}&=&C_{n+N}^+ \;. \label{tc+t}
\eean
It follows from (\ref{ta+t}), ({\ref{ta-t}), (\ref{ta0t}) and (\ref{tc-t}--\ref{tc+t})
that the action of the transformation $T^N$ on the Hamiltonian (\ref{hamiltonian})
is given by
\[
T^N \hat{H}(T^N)^\dag =\hat{H}-\frac{g^2 N \pi}{2L}(C_0^3 +D_0^3 ) +\frac{ig^2 N\pi}{2L}
\sum_{n=\half}^\infty \left( (a_n^+)^\dag b_n^+ + ( b_n^-)^\dag a_n^-  -(a_n^-)^\dag b_n^-
-( b_n^+)^\dag a_n^+ \right)\,. \\
\]
$\hat{H}$ is not invariant under the action of $T^N$ but
\[
T^N \hat{H}(T^N)^\dag =\hat{H}-\frac{gN \pi}{\sq}\lambda_0^3
\]
which means that $\hat{H}$ is invariant when restricted to the physical subspace.
Note also that
\[
T^N \lambda_0^3 (T^N)^\dag =\lambda_0^3
\]
so that  if $ \lambda_0^3 |\varphi\rangle =0$ then also $ \lambda_0^3 T^N |\varphi\rangle =0$ and
\[
\hat{H}T^N |\varphi\rangle = T^N\hat{H}(T^N)^\dag T^N  |\varphi\rangle
+\frac{g N \pi}{\sq}\lambda_0^3  T^N  |\varphi\rangle =T^N \hat{H}|\varphi\rangle\;.
 \]
In particular this means that the states
\[
|\Omega_N\rangle \equiv T^N |\Omega\rangle \ , \quad N=0,\pm1, \pm 2, \dots
\]
are an infinite set of degenerate vacua. These states are clearly
not gauge-invariant.  Physically acceptable gauge-invariant vacua
can be obtained by constructing superpositions that diagonalize
the operators $T^N$. We can create eigenstates of $T$ by forming $\theta$-states
as in the case of the Schwinger model.  If we take
\[
|\theta\rangle \equiv \sum_{N=-\infty}^\infty e^{-iN\theta}|\Omega_N \rangle \;,
\]
we have
\[
T^M |\theta\rangle =  \sum_{N=-\infty}^\infty e^{-iN\theta}|\Omega_{N+M} \rangle
=e^{iM\theta}|\theta\rangle
\]
so that $|\theta\rangle$ is invariant up to a phase factor under the action of $T^M$.

The theory is also invariant \cite{paper} under the transformation $R$ such that
\bea
&&R\psi R^{-1}=\psi^\dag \\
&&R\phi R^{-1}=-\phi \\
&&RA^\pm R^{-1}= A^\mp\\
&&RA^3 R^{-1}= -A^3
\eea
corresponding to the $SU(2)$ transformation $U$=\begin{large}$e^{i\pi\tau_1}$\end{large}.\\
The action of $R$ on the fermion Fock operators is
\bea
&&Rb_n R^{-1} =d_n \qquad   \\
&&R\beta_n R^{-1} =\delta_n \\
 &&Rr_N R^{-1} =-r_N   \qquad  \\
&&R\rho_N R^{-1} =-\rho_N \;.
\eea
As a consequence we have
\bea
&&R C_n^\pm R^{-1}= C_n^\mp  \\
&&RC_N^3 R^{-1}=-C_N^3 \;.
\eea
The gauge Fock operators transform as
\bea
&&RA_n^\pm R^{-1}= A_n^\mp\\
&&RA_N^3 R^{-1}= -A^3_N \;.
\eea
Note that the state $R \0$ is annihilated by all the destruction operators and, since $R^2=1$,
we must have
\[
R \0 = \pm \0 \;.
\]
Without loss of generality we may take $R \0=\0$. This relation, together with the transformation properties of the Fock operators, defines the action of $R$ on all states. \\
One can immediately see that the state $|\Omega\rangle$ is invariant under the action of $R$:
\[
R|\Omega\rangle =|\Omega\rangle \;.
\]
Let us consider now the action of $R$ on the other vacuum states $|\Omega_N \rangle\equiv T^N | \Omega\rangle$.
From the definition of the spurion operators (p. \pageref{sigmaRL}) it is not hard to
 see that
\[
R\sigma_{R/L}R^{-1}= \sigma_{R/L}^\dag
\]
and it is straightforward to verify that
\[
RTR^{-1}= -T^\dag
\]
and
\[
RT^N R^{-1}= (-1)^N (T^\dag)^N \equiv  (-1)^N T^{-N}\;.
\]
As a consequence, $R$ interchanges $|\Omega_N \rangle$ and $|\Omega_{-N} \rangle$
\[
R|\Omega_N \rangle =(-1)^N |\Omega_{-N} \rangle \;.
\]
By applying $R$ to the $\theta$-vacuum we obtain
\[
R|\theta\rangle = \sum_{N=-\infty}^{\infty}e^{-iN\theta}R|\Omega_{N} \rangle
= \sum_{N=-\infty}^{\infty}e^{-iN\theta}e^{iN\pi}|\Omega_{-N} \rangle
= \sum_{N=-\infty}^{\infty}e^{-iN(\pi-\theta)}|\Omega_{N} \rangle
\]
and we see that only two values of the parameter $\theta$ , namely $\theta=\pm \frac{\pi}{2}$, give rise to states which are invariant under both the $T$
and $R$ residual symmetries. We therefore have two physically acceptable vacua, in agreement
with  refs. \cite{witten}\cite{smilga}\cite{lenz}\cite{paper}.

\subsection{The Condensate}

We want to use our results for the vacuum to obtain a perturbative evaluation of the gauge-invariant  fermion condensate, defined as
\[
\frac{\langle\theta | {\rm Tr}\bar{\Psi}\Psi | \theta \rangle}{\langle\theta  | \theta \rangle} \;.
\]
We have
\[
{\rm Tr}\bar{\Psi}\Psi =i {\rm Tr }(\Psi_L^\dag \Psi_R-\Psi_R^\dag\Psi_L) =
i\left(\phi_L \phi_R + \psi_L^\dag \psi_R -  \psi_R^\dag \psi_L \right)
\]
and
\[
\langle\theta | {\rm Tr}\bar{\Psi}\Psi | \theta \rangle
=i\! \sum_{N,M=-\infty}^\infty e^{i (M-N)\theta}\, \langle \Omega_M |\left(\phi_L \phi_R + \psi_L^\dag \psi_R -  \psi_R^\dag \psi_L \right) |\Omega_N\rangle \;.
\]
Being a time-independent quantity, the fermion condensate can be evaluated at $t=0$. \\
Writing
\[
\Omega_N =\Omega_N^{(0)} +\Omega_N^{(1)} +\Omega_N^{(2)} +\ldots
\]
we have
\[
\langle\theta | {\rm Tr}\bar{\Psi}\Psi | \theta \rangle
=i \sum_{j,k} \sum_{N,M=-\infty}^\infty e^{i (M-N)\theta}\, \langle \Omega_M^{(j)} |\left(\phi_L \phi_R + \psi_L^\dag \psi_R -  \psi_R^\dag \psi_L \right)\!(0,\x)\;|\Omega_N^{(k)}\rangle
\]
where
\[
|\Omega^{(i)} _{N} \rangle= T^N |\Omega^{(i)} \rangle \quad {\rm and}
 \quad  | \Omega^{(0)} \rangle \equiv \0 \;.
\]
Explicitly
\[
| \Omega^{(0)}_N\rangle = T^N \0  = e^{-\frac{iN\pi\sq}{g\sqrt{L}}b_0^3} \left( \sigma_L^\dag \sigma_R\right)^N \0 \equiv
 e^{-\frac{iN\pi\sq}{g\sqrt{L}}b_0^3}|N\rangle
\]
where
\bea
&&|N\rangle =\beta_{N-\half}^\dag d_{N-\half}^\dag\cdots \beta_{\half}^\dag d_{\half}^\dag \0 \qquad
{\rm for} \ N>0\\
&&|N\rangle =\delta_{N-\half}^\dag b_{N-\half}^\dag\cdots \delta_{\half}^\dag b_{\half}^\dag \0 \qquad
\;{\rm for} \ N<0 \, ,
\eea
\bea
&&|\Omega^{(1)}_N\rangle =
-\frac{g \sq}{\sqrt{L}}\sum_{N=1}^{\infty}\left(\frac{{a_N^3}^\dag}{2k_N}+\frac{i{b_N^3}^\dag}{(2k_N)^2}
\right){C_N^3}^\dag  | \Omega^{(0)}_N\rangle \\
&&\hspace{25pt}-\frac{g}{\sqrt{mL}}
 \sum_{n=\half}^{\infty}\frac{1}{2k_n+m}\bigg( (A_{n+N}^-)^\dag (C_{n+N}^+)^\dag | \Omega^{(0)}_N\rangle  +(A_{n-N}^+)^\dag (C_{n-N}^-)^\dag | \Omega^{(0)}_N\rangle  \bigg)\, ,
\eea
\vskip0pt

\baselineskip15pt
\bea
|\Omega^{(2)}_N\rangle  \!\!\!&=&\!\!\!
\frac{g^2}{L}\, \sum_{M=1}^{\infty}\sum_{J=1}^{\infty}
\left(\frac{{a_M^3}^\dag}{2k_M}+\frac{i{b_M^3}^\dag}{(2k_M)^2}
\right)\left(\frac{{a_J^3}^\dag}{2k_J}+\frac{i{b_J^3}^\dag}{(2k_J)^2}
\right){C_M^3}^\dag {C_J^3}^\dag  | \Omega^{(0)}_N\rangle \\ \\
&\;+&\!\!\!\! \frac{g^2 \sq}{\sqrt{m}L}\sum_{M=1}^{\infty}\sum_{n=\half}^{\infty}\Bigg(\frac{{a_M^3}^\dag}{2k_{M}+2k_n +m}\\
&&\hspace{30pt}+\frac{i{b_M^3}^\dag}{(2k_{n}+2k_M +m)^2} \Bigg){C_M^3}^\dag \, \frac{ (A_{n+N}^-)^\dag  (C_{n+N}^+)^\dag +(A_{n-N}^+)^\dag (C_{n-N}^-)^\dag }{2k_n+m}| \Omega^{(0)}_N\rangle \\ \\
 &\;+&\!\!\!\! \frac{g^2 \sq}{\sqrt{m}L}\sum_{J=1}^{\infty}\sum_{p=-\infty}^{\infty}
\Bigg\{ \frac{1}{2k_{p}+2k_J +m} \left(\frac{{a_J^3}^\dag}{2k_J}+\frac{i{b_J^3}^\dag}{(2k_J)^2}
\right) \\
&&\hspace{15pt}+ \frac{i{b_J^3}^\dag}{2k_J (2k_{p}+2k_J +m)^2}\Bigg\}
\bigg((A_{p+N}^-)^\dag   (C_{p+N}^+)^\dag +(A_{p-N}^+)^\dag (C_{p-N}^-)^\dag  \bigg){C_J^3}^\dag
| \Omega^{(0)}_N\rangle \\ \\
&\;+&\!\!\!\! \frac{g^2}{mL}\Bigg\{ \sum_{n=\half}^{\infty} \frac{\left( (A_0^3)^\dag -\frac{N\pi \sqrt{m}}{g\sqrt{L}}\right)
 \bigg({ -(A_{n+N}^-)^\dag  (C_{n+N}^+)^\dag +(A_{n-N}^+)^\dag (C_{n-N}^-)^\dag }\bigg)}
{2(k_{n}+m) (2k_n+m)}| \Omega^{(0)}_N\rangle\\ \\
&& \hspace{20pt}+\sum_{p=\half}^{\infty}\sum_{n=\half}^{\infty} \frac{(A_{p-N}^+)^\dag (A_{n-N}^+)^\dag(C_{p-N}^-)^\dag   (C_{n-N}^-)^\dag  }{2(2k_p +m)(2k_n+m)}| \Omega^{(0)}_N\rangle\\ \\
&& \hspace{20pt}+ \sum_{p=\half}^{\infty}\sum_{n=\half}^{\infty}+\frac{(A_{p+N}^-)^\dag (A_{n+N}^-)^\dag(C_{p+N}^+)^\dag   (C_{n+N}^+)^\dag  }{2(2k_p +m)(2k_n+m)}|\Omega^{(0)}_N\rangle\\ \\
&& \hspace{20pt}+\sum_{p=-\infty}^{\infty}\sum_{n=\half}^{\infty} \frac{(A_{p+N}^-)^\dag (A_{n-N}^+)^\dag(C_{p+N}^+)^\dag   (C_{n-N}^-)^\dag  }{2(k_{n}+k_p +m)(2k_n+m)}
| \Omega^{(0)}_N\rangle\\ \\
&& \hspace{20pt}+\sum_{p=-\infty}^{\infty}\sum_{n=\half}^{\infty}
\frac{(A_{p-N}^+)^\dag (A_{n+N}^-)^\dag(C_{p-N}^-)^\dag   (C_{n+N}^+)^\dag  }{2(k_{n}+k_p +m)(2k_n+m)}|\Omega^{(0)}_N\rangle
 \Bigg\}.
\eea
We define
\[
|\theta ^{(i)}\rangle = \sum_{N=-\infty}^{\infty}e^{-iN\theta}|\Omega^{(i)} _{N} \rangle
\]
so that we can write
\[
|\theta\rangle= |\theta ^{(0)}\rangle +|\theta ^{(1)}\rangle + |\theta ^{(2)}\rangle +\ldots
\]
and
\[
 \langle\theta | {\rm Tr}\bar{\Psi}\Psi | \theta \rangle=\sum_{i,j}
\langle\theta^{(i)} | {\rm Tr}\bar{\Psi}\Psi | \theta^{(j)} \rangle
\]

The calculation of this quantity is long and tedious, involving very lengthy expressions.  Here, 
we shall just summerize a few of the intermediate steps and give the result.  We first consider the 
contribution of the complex field.  We find that
\bea
\langle \theta |i( \psi_L^\dag \psi_R  - \psi_R^\dag \psi_L )  | \theta \rangle &  \simeq &
 \langle \theta^{(0)}| i(\psi_L^\dag  \psi_R -\psi_R^\dag \psi_L )   | \theta^{(0)}\rangle
+\langle \theta^{(1)}| i(\psi_L^\dag  \psi_R-\psi_R^\dag \psi_L )   | \theta^{(1)}\rangle \\
&&\!\!+2 \langle \theta^{(1)}| i(\psi_L^\dag  \psi_R-\psi_R^\dag \psi_L )   | \theta^{(2)}\rangle
+\langle \theta^{(2)}| i( \psi_L^\dag  \psi_R-\psi_R^\dag \psi_L  ) | \theta^{(2)}\rangle /; ,
\eea
where
\be  \label{condzero}
\langle \theta^{(0)}|i\left(\psi_L^\dag (0,\x) \psi_R (0,\x) -\psi_R^\dag (0,\x) \psi_L (0,\x)\right)| \theta^{(0)}\rangle=
 -\frac{1}{L}
\sum_{N=-\infty}^\infty \sin \theta e^{-\frac{\pi}{2mL}}  \; ,
\ee
\bean
&&\hspace{-50pt} \langle \theta^{(1)}|\,i\left( \psi_L^\dag (0,\x) \psi_R (0,\x)- \psi_L^\dag (0,\x) \psi_R (0,\x)\right) | \theta^{(1)}\rangle = \nn \\ \nn  \\
&&\hspace{-30pt}-\frac{g^2}{mL^2}e^{-\frac{\pi}{2mL}}\sum_{N=-\infty}^{\infty}\sin \theta\Bigg\{
 \sum_{n=\half}^{\infty}\frac{2 n}{(2k_{n}+m)(2k_{n+1}+m)}-m\sum_{J=1}^{\infty}\frac{(J-1)}{2k_J^3}\Bigg\}\,, \label{condone} \\ \nn
\eean
\bean
&&\langle \theta^{(1)}|\,i\left( \psi_L^\dag (0,\x) \psi_R (0,\x)- \psi_R^\dag (0,\x) \psi_L (0,\x)\right) | \theta^{(2)}\rangle=  \nn \\
&&= \frac{g^2\pi^2 }{mL^4}e^{-\frac{\pi}{2mL}}\sum_{N=-\infty}^{\infty}\sin \theta
\sum_{n=\half}^{\infty}\frac{n}{2(2k_{n}+m)(2k_{n+1}+m)(k_{n}+m)(k_{n+1}+m)} \;, \label{condonetwo} \nn \\
\eean
and
\bea
&&\langle \theta^{(2)}| i\psi_L^\dag  \psir-i\psi_R^\dag \psi_L   | \theta^{(2)}\rangle= \\ \\
&&\quad=-\frac{g^4e^{-\frac{\pi}{2mL}}}{m^2 L^3}\sum_{N=-\infty}^{\infty}\sin \theta
\Bigg\{
\sum_{I,J=1}^{\infty}
\frac{8m^2}{(2k_I)^3 ((2k_J)^3}(IJ-I-J+1)\\
&&\hspace{35pt}+ \sum_{J=1}^{\infty}
\frac{8m^2}{(2k_J)^6}(J^2-2J) \\ \\
&&\hspace{35pt}-8m \sum_{J=1}^{\infty}
\sum_{p=\half}^{\infty}\frac{pJ-J+1-p +\theta(J-p-1) (J-p-1)}
{(2k_p +m) (2k_{p+1} +m)(2k_J)^3}\\ \\
&&\hspace{35pt} -4m\sum_{J=1}^{\infty}\sum_{p=\half}^{\infty} \frac{p}{(2k_p+m)(2k_{p+1}+m)}
\Bigg(\frac{1}{(2k_p +2k_J +m)^2 (2k_{p+1} +2k_J +m)}\\
&&\hspace{205pt}+\frac{1}{(2k_p +2k_J +m) (2k_{p+1} +2k_J +m)^2}\Bigg)\\ \\
&&\hspace{35pt}\!+2m \sum_{J=1}^{\infty}\sum_{p=\half}^{\infty}\! \frac{1 }{(2k_{p+1}+m)}
\Bigg(\! \frac{2}{(2k_J)^3 (2k_p +2k_J +m)}\\
&&\hspace{165pt}+\frac{1}{(2k_J)^2 (2k_p +2k_J +m)^2}\!\Bigg) \\ \\
&&\hspace{35pt}-4m\sum_{J=1}^{\infty}\sum_{p=\half}^{\infty}
\Bigg(\frac{2J}{(2k_J)^3 (2k_p +2k_J +m)(2k_{p+1} +2k_J +m)}\\
&&\hspace{85pt}+\frac{J}{(2k_J)^2 (2k_p +2k_J +m)^2
 (2k_{p+1} +2k_J +m)}\\
&&\hspace{85pt}+ \frac{J}{(2k_J)^2 (2k_p +2k_J +m) (2k_{p+1} +2k_J +m)^2}\Bigg)\\ \\
&&\hspace{35pt}+\sum_{J=1}^{\infty}\frac{m }
{(2k_{\half}+m)}\Bigg(\frac{2}{(2k_J)^3 (2k_{J-\half} +m)}+\frac{1}{(2k_J)^2 (2k_{J+\half} +m)^2}\Bigg) \\ \\
&&\hspace{35pt}-4m \sum_{J=1}^{\infty}\sum_{p=\half}^{\infty}
\Bigg(\frac{2\theta(J-p)(J-p)}{(2k_J)^3 ( 2k_{J-p} +m)(2k_{J-p+1} +m)}\\
&&\hspace{85pt}+\frac{\theta(J-p)(J-p)}{(2k_J)^2 (2k_{J-p} +m)^2
 (2k_{J-p+1} +m)}\\
&&\hspace{85pt}+ \frac{\theta(J-p)(J-p)}{(2k_J)^2 (2k_{J-p} +m) (2k_{J-p+1} +m)^2}\Bigg)
\\ \\
&& \hspace{35pt}+ \sum_{p=\half}^{\infty} \frac{\left(1-\frac{\pi^2 m}{g^2 L}\right)p
}{2(k_{p}+m) (2k_p+m)(k_{{p+1}}+m) (2k_{p+1}+m)}\\ \\
&&\hspace{35pt}+\sum_{p,n=\half}^{\infty}\frac{2np-n+(n-p)\theta(n-p -\half)}
{(2k_p  +m) (2k_n +m) (2k_{p+1} +m) (2k_{n+1} +m)}\\ \\
&&\hspace{35pt}+\sum_{n=\half}^{\infty}\frac{n^2}{(2k_n  +m)^2 (2k_{n+1} +m)^2 }\\ \\
&&\hspace{40pt}+ \sum_{n=\half}^{\infty}\sum_{p=\half}^{\infty}
\frac{n}
{2 (2k_n +m)  (2k_{n+1} +m)(k_{n+p+1} +m)^2}
\\ \\
&&\hspace{40pt}+\sum_{p=\half}^{\infty}\sum_{n=\half}^{\infty}\frac{(n-p-1)\theta(n-p-\half)}
{2(2k_n +m)  (2k_{n+1} +m)(k_{n-p} +m)^2}\\ \\
&&\hspace{40pt}+\sum_{n=\half}^{\infty}\frac{n-\half}
{2(2k_n +m)  (2k_{n+1} +m)(k_{n+\half} +m)^2}\\ \\
&&\hspace{40pt}+ \frac{1}
{16  (2k_{\half} +m)^2  m^2}
+\sum_{n=\half}^{\infty}\frac{n(n+1) }
{2 (2k_n +m)  (2k_{n+1} +m) m^2}\, \Bigg\}
\:.
\\
\eea
We shall also need the norm of the state.  We have that
\bea
\langle \theta  | \theta \rangle &\simeq &
\langle \theta^{(0)}|  \theta^{(0)}\rangle
+\langle \theta^{(1)}|  \theta^{(1)}\rangle
+\langle \theta^{(2)}|  \theta^{(2)}\rangle \; ,
\eea
where
\be
 \label{normzero} \langle \theta ^{(0)}|
\theta^{(0)}\rangle= \sum_{N,M=-\infty}^\infty e^{i (M-N)\theta}\,
\langle \Omega_M^{(0)}| \Omega^{(0)}_N\rangle =
\sum_{N,M=-\infty}^\infty e^{i (M-N)\theta}\delta_{M,N}=
\sum_{N=-\infty}^\infty 1  \; ,
\ee
\be \label{normone}
 \langle \theta^{(1)}| \theta^{(1)}\rangle=\frac{g^2}{mL}\sum_{N=-\infty}^{\infty}\Bigg\{-\sum_{J=1}^{\infty} \frac{mJ}{2k_J^3}
+\sum_{n=\half}^{\infty}\frac{2 n}{(2k_{n}+m)^2}\Bigg\} \; ,
\ee
and
\bea
&&\langle \theta^{(2)}|  \theta^{(2)}\rangle =
\frac{g^4}{m^2 L^2}\sum_{n=-\infty }^{\infty}\Bigg\{
\sum_{I,J=1}^{\infty}IJ \frac{8m^2}{(2k_I)^3 ((2k_J)^3}
+\sum_{J=1}^{\infty}\frac{8m^2}{(2k_J)^6}J^2
 \\ \\
&&\hspace{50pt} -8m\sum_{p=\half}^{\infty}\sum_{J=1}^{\infty} \frac{pJ-p +\theta (p-J) (p-J)}{(2k_p+m)^2 (2k_J)^3}\\ \\
&& \hspace{50pt}
-8m \sum_{p=\half}^{\infty}\sum_{J=1}^{\infty}
\frac{p}{(2k_p+m)^2  (2k_p +2k_J + m )^3}\\ \\
&&\hspace{50pt}-8m\sum_{p=\half}^{\infty}\sum_{J=1}^{\infty}
\Bigg( \frac{J}
{(2k_p +2k_J+m )^2 (2k_J )^3 }+\frac{J}{(2k_p +2k_J +m)^3 (2k_J)^2 }
\Bigg)\\ \\
&&\hspace{50pt}-8m\sum_{p=\half}^{\infty}\sum_{J=1}^{\infty}
\Bigg( \frac{\theta(J-p) (J-p)}
{(-2k_p +2k_J+m )^2 (2k_J )^3 }+\frac{\theta(J-p) (J-p)}{(-2k_p +2k_J +m)^3 (2k_J)^2 }
\Bigg)\\ \\
&& \hspace{50pt}+ \sum_{p=\half}^{\infty} \frac{p}{2(k_{p}+m)^2 (2k_p+m)^2}
 +\sum_{n=\half}^{\infty} \frac{n^2}{(2k_n +m)^4}\\ \\
&& \hspace{50pt}+\sum_{p,n=\half}^{\infty}\frac{2np-n
+(n-p)\theta(n-p-\half)}{(2k_p  +m)^2 (2k_n +m)^2}
+ \sum_{n=\half}^{\infty} \!\frac{ n^2}
{2m^2 (2k_n +m)^2 }\\ \\
&& \hspace{50pt}+ \sum_{n, p=\half}^{\infty}\Bigg[
 \!\frac{(n-p)\theta(n-p-\half)}
{2(k_{n-p} +m)^2 (2k_n +m)^2 }
+\frac{n}
{2(k_{p +n} +m)^2 (2k_n +m)^2  }\Bigg]
\Bigg\}\;.
\eea
Using these results we can calculate $\frac{\langle \theta | \psi_L^\dag \psi_R | \theta \rangle}
{\langle \theta  | \theta \rangle }$:
\bea
&&\hspace{-20pt}\frac{ \langle \theta | i(\psi_L^\dag \psi_R - \psi_R^\dag \psi_L )| \theta \rangle}
{ \langle \theta  | \theta \rangle }\simeq \\ \\
&&\hspace{-20pt}\simeq \sin\theta e^{-\frac{\pi}{2mL}}\Bigg\{
-\frac{1}{L}+\frac{g^2}{mL^2}\Bigg[  \sum_{n=\half}^{\infty}\frac{4k_n}{(2k_n +m)^2 (2k_{n+1} +m)}
-\sum_{J=1}^{\infty}\frac{4m}{(2k_J )^3}\Bigg] \\ \\
&&\hspace{-15pt}+\frac{g^4}{m^2 L^3}\Bigg[ - \sum_{n=\half}^{\infty}\frac{k_n }{(2k_n +m)^2 (2k_{n+1} +m)}
\sum_{p=\half}^{\infty}\frac{4 k_p}{ (2k_p +m)^2 (2k_{p+1} +m)}\\ \\
&&\hspace{20pt}+\sum_{n=\half}^{\infty}\frac{4k_n}{(2k_n +m)^2 (2k_{n+1} +m)}\sum_{J=1}^{\infty}
\frac{4m}{(2k_J )^3}\\ \\
&&\hspace{20pt}+\sum_{J=1}^{\infty}
\frac{4mJ}{(2k_J )^3}
\sum_{n=\half}^{\infty}\frac{4k_n}{(2k_n +m)^2 (2k_{n+1} +m)}-\sum_{I,J=1}^{\infty}
\frac{8m^2}{(2k_I )^3 (2k_J)^3}
+\sum_{J=1}^{\infty}\frac{16m^2 J}{ (2k_J)^6}\\ \\
&&\hspace{20pt}+\sum_{p=\half}^{\infty}
\sum_{J=1}^{\infty}
\frac{4mp}{(2k_p+m) (2k_{p} +2k_J+m )}\Bigg(\frac{1}{(2k_{p+1} +m )(2k_{p+1} +2k_J+m )^2 }\\
&&\hspace{20pt}
+\frac{1}{(2k_{p+1} +m )(2k_{p+1} +2k_J+m ) (2k_{p} +2k_J+m )}\\
&&\hspace{20pt}-\frac{2}{(2k_p+m) (2k_{p} +2k_J+m )^2} \Bigg)\\ \\
&&\hspace{20pt}-\sum_{J=1}^{\infty}\sum_{p=\half}^{\infty}\frac{ 2m}
{(2k_{p+1}+m)(2k_p +2k_J +m)(2k_J)^2}\Bigg( \frac{1}{(2k_p +2k_J +m)}+ \frac{1}{k_J}
\Bigg)\\ \\
&&\hspace{20pt}-\sum_{p=\half}^{\infty}\sum_{J=1}^{\infty} \frac{8mk_J}
{(2k_p +2k_J +m) (2k_J)^2 }\Bigg(\frac{1}{k_J (2k_p +2k_J +m)(2k_{p+1} +2k_J +m)}\\
&&\hspace{20pt}\!+\frac{1}{(2k_p +2k_J +m)(2k_{p+1} +2k_J +m)^2}
+\! \frac{2}{(2k_p +2k_J +m)^2(2k_{p+1} +2k_J +m)}\Bigg)\\ \\
&&\hspace{20pt}+\sum_{J=1}^{\infty}\frac{ m}
{(2k_{\half}+m)(2k_{J-\half} +m)(2k_J)^2}\Bigg( \frac{1}{(2k_{J-\half} +m)}+ \frac{1}{k_J}
\Bigg)\\ \\
&&\hspace{20pt}-\pl \sum_{p=\half}^{\infty}\sum_{J=1}^{\infty} \frac{8m\theta(J-p) (J-p)}
{(-2k_p +2k_J +m) (2k_J)^2 }\Bigg(\frac{1}{(-2k_p +2k_J +m)(-2k_{p-1} +2k_J +m)^2}\\
&&\hspace{20pt}+\frac{1}{k_J (-2k_p +2k_J +m)(-2k_{p-1} +2k_J +m)}\\
&&\hspace{20pt}+\! \frac{2}{(-2k_p +2k_J +m)^2(-2k_{p-1} +2k_J +m)}\Bigg)\\ \\
&&\hspace{20pt}-  \sum_{p=\half}^{\infty}\sum_{J=1}^{\infty}\frac{16m}{(2k_J)^3}
 \frac{k_p J -k_J +\theta (J-p) (k_J -k_p )}{(2k_p+m)^2 (2k_{p+1} +m)}\\ \\
&&\hspace{20pt}+\sum_{p=\half}^{\infty}\sum_{J=1}^{\infty}\frac{8m}{(2k_J)^3}
\frac{1 -\theta(J-p-1) }{(2k_p +m) (2k_{p+1} +m)}\\ \\
&&\hspace{20pt}-\sum_{p=\half}^{\infty}\frac{4m}{(2k_{p+\half})^3(2k_p +m) (2k_{p+1} +m)}\\ \
&&\hspace{20pt}+\sum_{p=\half}^{\infty} \frac{k_p}{2(k_{p}+m)^2 (2k_p+m) (2k_{p+1}+m)}
\Bigg( \frac{2}{(2k_p+m)}+\frac{1}{(k_{p+1}+m)}\Bigg)\\ \\
&&\hspace{20pt}+\frac{3\pi^2 m}{2g^2 L}\sum_{p=\half}^{\infty} \frac{p
}{(k_{p}+m) (2k_p+m)(k_{{p+1}}+m) (2k_{p+1}+m)}\\ \\
&&\hspace{20pt}+ \sum_{p,n=\half}^{\infty}\frac{-2k_n \theta(p-n+\half)-2 k_p \theta(n-p-\half) }
{(2k_p  +m) (2k_n +m)^2 (2k_{p+1} +m)}\Bigg( \frac{ 1}{(2k_p  +m) }+\frac{1}{  (2k_{n+1} +m)} \Bigg)  \\ \\
&&\hspace{20pt}+\sum_{n=\half}^{\infty} \frac{n k_n}{(2k_n +m)^3(2k_{n+1} +m)}
\Bigg(\frac{1}{(2k_{n+1} +m) }+
\frac{1}{(2k_n  +m) }
\Bigg) \\ \\
&&\hspace{20pt}-\sum_{n=\half}^{\infty} \!\frac{ n }
{2m (2k_n +m)^2 (2k_{n+1} +m) }\\ \\
&& \hspace{20pt}+\sum_{n, p=\half}^{\infty}
 \!\frac{(k_n-k_p)\theta(n-p-\half)}
{(k_{n-p} +m)^2 (2k_n +m)^2 (2k_{n+1} +m)}\\ \\
&&\hspace{20pt}+\sum_{p=\half}^{\infty}\sum_{n=\half}^{\infty}\frac{\theta(n-p-\half)}
{2(2k_n +m)  (2k_{n+1} +m)(k_{n-p} +m)^2}\\ \\
&&\hspace{20pt}+\sum_{n=\half}^{\infty}\frac{n-\half}
{2(2k_n +m)  (2k_{n+1} +m)(k_{n+\half} +m)^2}
+ \frac{1}{16  (2k_{\half} +m)^2  m^2}\\ \\
&&\hspace{20pt}+ \sum_{n=\half}^{\infty}\sum_{p=\half}^{\infty}
\frac{k_n}
{2 (2k_n +m ) (2k_{n+1} +m)(k_{p +n} +m)}\Bigg(\frac{1}{(k_{p +n} +m) (k_{p +n+1} +m)}\\ \\
&&\hspace{165pt}+\frac{1}{ (k_{p +n+1} +m)^2 }
+ \frac{2}{(k_{p +n} +m) (2k_n +m )}\Bigg)\Bigg] \Bigg\}
\eea
While some of the sums in $\langle \theta |i( \psi_L^\dag \psi_R  - \psi_R^\dag \psi_L )  | \theta \rangle$ 
and $\langle \theta  | \theta \rangle$ are divergent, the ratio is finite.

We also need to consider the potential contribution of the real field.  We find that
\bea
&&\frac{\langle \theta | \phi_L (0,\x) \phi_R (0,\x) |\theta \rangle}{\langle \theta | \theta \rangle}\simeq
\langle 0| { \stackrel {\;o}{\phi}}_L  {\stackrel {\;o}{\phi}}_R \0  \Bigg\{ 1 -2 \frac{g^2}{mL}  \sum_{n=\half}^{\infty}\frac{1}{(2k_n+m)^2} \\ \\
&&\hspace{40pt}+
 \frac{g^4}{m^2 L^2}  \Bigg[
8m\sum_{p=\half}^{\infty} \sum_{J=1}^{\infty}\frac{1 }{(2k_p+m)^2 (2k_p +2k_J + m )^3}\\ \\
&&\hspace{60pt}+8m\sum_{p=\half}^{\infty}\sum_{J=1}^{\infty}\frac{1}{(2k_{J+p-\half} )^2 (2k_{J-\half} +m )^3 }
\\ \\
&& \hspace{60pt}- \sum_{p=\half}^{\infty}\frac{1}{4(k_{p}+m)^2 (2k_p+m)^2}-\sum_{p,n=\half}^{\infty}\! \frac{-2 +2p\delta_{n,p}}
{(2k_p  +m)^2 (2k_n +m)^2}\Bigg] \Bigg\}
\eea
and it is not hard to verify that it goes to zero in large-L limit,  as one might have expected considering
that this contribution is an artifact of the breaking of chiral invariance in free theory due to the zero 
modes.

We now want to evaluate the condensate in the large $L$ limit.  By studying the large-$L$  behaviour  one can see
that, while several terms go to zero, others  diverge with $L$. The divergent behaviour is expected since it 
is found also in the expansion of the factor multiplying the exponential $e^{-\frac{\pi}{2mL}}$ in the 
finite-L condensate for the Schwinger model. In that case, the full nonperturbative result has a finite 
limit \cite{eliana}.

Setting $g=m\sqrt{\pi}$ we can see that the condensate takes the form (recall that $\theta = \pm \frac{\pi}{2}$)
\be
 \frac{ \langle \theta | i(\psi_L^\dag \psi_R - \psi_R^\dag \psi_L )| \theta \rangle}
{ \langle \theta  | \theta \rangle } = m \sin\theta e^{-\frac{\pi}{2mL}} f(m L)
\ee
 where $f$ is a function of $m L$ that goes
to a pure number, if convergent, in the large-$L$ limit . We therefore find that, as in the case of
the Schwinger model, the condensate is proportional to
the coupling constant. 

Disregarding contributions that vanish as L goes to infinity we get the following estimate for the large-L behaviour of the condensate
\begin{eqnarray*}
\frac{ \langle \theta | i(\psi_L^\dag \psi_R - \psi_R^\dag \psi_L )| \theta \rangle}
{ \langle \theta  | \theta \rangle }&\simeq&  m \sin\theta e^{-\frac{\pi}{2mL}}
 \Bigg( \frac{1}{12 \pi}\sum_{J}\frac{1}{J^3} 
 - \frac{7 }{8  \pi}\sum_{J}\frac{1}{J^2}\\ 
&&- \frac{m^3 L^3}{8  \pi^4}\left(\sum_{J}\frac{1}{J^3}\right)^2
+ \frac{m^3 L^3}{4  \pi^4}\sum_{J}\frac{1}{J^5}\Bigg)
\end{eqnarray*}
A standard technique to estimate the value of a function in the limit where the argument goes 
to infinity when the function is defined by a power series is that of  Pad\'e approximants\cite{gary}. 
The method of quadratic approximants gives the following function
\[
f(x) = \frac{-1 +\sqrt{1+4 x^3 (a+(a^2 +b)x^3 )}}{2 x ^3}
\]
which has the  power series expansion
\[
f(x) \simeq a + b x^3 \;.
\] 
For our case
 \bea
&&a= \frac{1}{12 \pi}\sum_{J}\frac{1}{J^3} 
 - \frac{7 }{8  \pi}\sum_{J}\frac{1}{J^2}
\simeq -0.426 \\
&& b=- \frac{1}{8  \pi^4}\left(\sum_{J}\frac{1}{J^3}\right)^2
+ \frac{1}{4  \pi^4}\sum_{J}\frac{1}{J^5}\simeq 0.000807\\
&& x=mL\,.
\eea
For $0< x < \infty$  $f$ is between $-0.426$ and $0.427$. It is therefore likely that the number 
multiplying $m\sin\theta$ in the condensate is within this range. But nothing definite can be 
said about the accuracy of this result. Since we can only form one approximant, we cannot test 
for convergence, even empirically, and there is no mathematical theorem giving a bound on the 
error that is made with this approximation. A similar procedure for the Schwinger model gives a 
correct estimate of the order of magnitude, with an asymptotic value of about 0.45, while the 
correct value is about 0.28. In all likelihood our number is of order 1. 

\section{Summary}

In quantizing $QCD_{1+1}$ with quarks in the adjoint representation with twisted boundary conditions 
we have encountered the unexpected difficulty that if all the components of the gauge field are 
quantized in indefinite metric (the quantization procedure we expected to have to use and the one 
that is required in the case of the Schwinger model), the residual gauge symmetry present at the 
classical level cannot be implemented at the quantum level.  On the other hand, if we quantize all the 
components of the gauge field in positive metric we find that, as expected, we cannot consistently 
remove the unphysical states from the system.  Our solution to this quandary relies explicitly on the 
twisted boundary conditions: we quantize the periodic gauge field in indefinite metric and the 
antiperiodic components of the gauge field in positive metric.  It is an open question as to what 
procedure should be used in the continuum or in a case where each of the components of the gauge 
field is subject to the same periodicity conditions.  The problem may have some importance since the 
same issue arises in the light-cone quantization of standard QCD.

With that mixed quantization scheme in place we 
show that a physical subspace exists which consists of color singlets .  We explicitly demonstrate 
that the physical subspace is dynamically stable by 
showing that the Lagrange multiplier fields are free fields.  We give the algebra of the Lagrange multiplier 
fields, which is likely to be of the same form as that in a continuum solution.  We show that the use of 
the mixed quantization scheme allows the quantum implementation of the residual gauge symmetry which 
exists at the classical level.  We then showed that there are two possible vacua, in agreement with the results 
of \cite{witten}\cite{smilga}\cite{lenz}\cite{paper}.  

We worked out the Hamiltonian.  If the Hamiltonian is partitioned into the kinetic energies, as the unperturbed 
Hamiltonian, and the interaction, as the perturbing Hamiltonian, perturbation theory cannot be applied.  
That is to be expected since the condensate, and the vacuum which leads to it, are nonperturbative 
quantities in the usual meaning of the word.  But we find that if we include a small part of the interaction 
in the unperturbed Hamiltonian we can diagonalize this new unperturbed Hamiltonian by hand and in doing 
so we include all the singular structure in these analytic unperturbed eigenstates.  We can now apply standard 
perturbation theory using the rest of the interaction as the perturbing operator.  We applied that perturbation 
theory to work out an expansion for the vacuum through the first three orders (two orders in the 
perturbation).  We then showed that the resulting state satisfies, through the relevant order, the subsidiary 
condition and is therefore a physical state.  With that vacuum we then evaluated, through the same order in 
perturbation theory, the chiral condensate.  We were able to show the dependence of the condensate on the 
parameters but have only an approximate value for a constant of proportionality.  We used Pad\'e approximants 
to find an approximate value for the constant in the limit where $L \rightarrow \infty$ (the so called, 
decompactification limit).

Several extensions of the work are possible.  Perhaps the most interesting, and possibly important, is to study 
the continuum case or a case where all components of the gauge field are periodic (untwisted boundary 
conditions) and determine what quantization procedure will allow the implementation of the residual gauge 
symmetry and a consistent implementation of the dynamics.  The perturbation theory could be used to find 
approximation to states other than the vacuum such as the one particle states (and thus determine an 
approximate spectrum).  It would be interesting to compare such results, and also the results we have presented 
here, with results from other techniques if such results become available.  It is possible that, if the 
calculations are automated on a computer, higher orders of perturbation theory could be worked out, thus 
allowing better estimates of the values and of the accuracy of the values.  We do not think it practical 
to try to go to higher order by hand.

\section*{Acknowledgments}

This work was supported by the Department of Energy through
contract DE-FG03-95ER40908 (G.M.).


\end{document}